\documentclass[transmag]{IEEEtran}
\usepackage{latexsym}
\usepackage{graphicx}
\usepackage{amsfonts,amssymb,amsmath}
\usepackage{hyperref,url}
\usepackage{color,array,hhline,multirow,booktabs}
\usepackage{subcaption,bigstrut}
\usepackage{epstopdf,url,hhline}
\usepackage[noadjust]{cite}
\usepackage[flushleft]{threeparttable}
\def\BibTeX{{\rm B\kern-.05em{\sc i\kern-.025em b}\kern-.08em T\kern-.1667em\lower.7ex\hbox{E}\kern-.125emX}}

\newtheorem{lemma}{Lemma}

\newcommand{\NRF}{N_\mathrm{RF}^\mathrm{t}}
\newcommand{\Nt}{N_\mathrm{t}}
\newcommand{\FRF}{\mathbf{F}_\mathrm{RF}}
\newcommand{\FBB}{\mathbf{F}_\mathrm{BB}}
\newcommand{\Fopt}{\mathbf{F}_\mathrm{opt}}

\newcommand{\invifont}{\fontsize{11bp}{13.5bp}\selectfont}

\begin{document}

\title{Hybrid Beamforming for 5G and Beyond Millimeter-Wave Systems: A Holistic View}

\author{Jun Zhang,
\IEEEmembership{Senior Member, IEEE},
Xianghao Yu,
\IEEEmembership{Member, IEEE}, and
Khaled B. Letaief,
\IEEEmembership{Fellow, IEEE}
\thanks{
	This work was supported by the Hong Kong Research Grants Council under Grant No. 16210216.}
\thanks{J. Zhang is with the Department of Electronic and Information Engineering (EIE), The Hong Kong Polytechnic University (PolyU) (e-mail: jun-eie.zhang@polyu.edu.hk).}

\thanks{X. Yu is with the Institute for Digital Communications, Friedrich-Alexander-University Erlangen-Nurnberg, 91054 Erlangen, Germany (e-mail: xianghao.yu@fau.de).
The work of X. Yu was supported by the Alexander von Humboldt
Foundation.}

\thanks{K. B. Letaief is with the Department of Electronic and Computer Engineering,
	The Hong Kong University of Science and Technology (HKUST) (e-mail:eekhaled@ust.hk), and also with
Peng Cheng Laboratory.}
}

\IEEEtitleabstractindextext{
\begin{center}
		{\invifont\vspace{-0.3em}\emph{(Invited Paper)}\vspace{1.3em}}
\end{center}

\begin{abstract}
Millimeter-wave (mm-wave) communication is a key technology for future wireless networks. To combat significant path loss and exploit the abundant mm-wave spectrum, effective beamforming is crucial. Nevertheless, conventional fully digital beamforming techniques are inapplicable, as they demand a separate radio frequency (RF) chain for each antenna element, which is costly and consumes too much energy. Hybrid beamforming is a cost-effective alternative, which can significantly reduce the hardware cost and power consumption by employing a small number of RF chains. This paper presents a holistic view on hybrid beamforming for 5G and beyond mm-wave systems, based on a new taxonomy for different hardware structures. We take a pragmatic approach and compare different proposals from three key aspects: 1) \emph{hardware efficiency}, i.e., the required hardware components; 2) \emph{computational efficiency} of the associated beamforming algorithm; and 3) achievable \emph{spectral efficiency}, a main performance indicator. Through systematic comparisons, the interplay and trade-off among these three design aspects are demonstrated, and promising candidates for hybrid beamforming in future wireless networks are identified.
\end{abstract}

\begin{IEEEkeywords}
Hybrid beamforming, millimeter-wave communications, 5G and beyond, wireless communications.
\end{IEEEkeywords}
}


\maketitle

\section{Introduction}
\IEEEPARstart{T}{he} continued upsurge of mobile data and the eruption of diversified mobile applications are driving the demand for next-generation wireless networks, i.e., the fifth-generation (5G) networks. Compared with the current fourth-generation (4G) Long Term Evolution (LTE) networks \cite{junLTE}, 5G needs to achieve orders of magnitude increase in the peak data rate, area spectral efficiency, network energy efficiency, while supporting a roundtrip latency of about 1 ms \cite{6824752}. Thus, disruptive technologies will be needed, and deploying 5G systems at millimeter wave (mm-wave) bands has been proposed due to the abundant spectrum. Thanks to the small wavelength of the mm-wave signals, large-scale antenna arrays can be deployed, and recent advances in massive multiple-input multiple-output (MIMO) \cite{6736761} can be leveraged
to provide beamforming gains to combat the increased path loss and synthesize highly directional beams to support mm-wave communications \cite{6515173}.

\begin{figure*}[t]
	\centering\includegraphics[width=17.5cm]{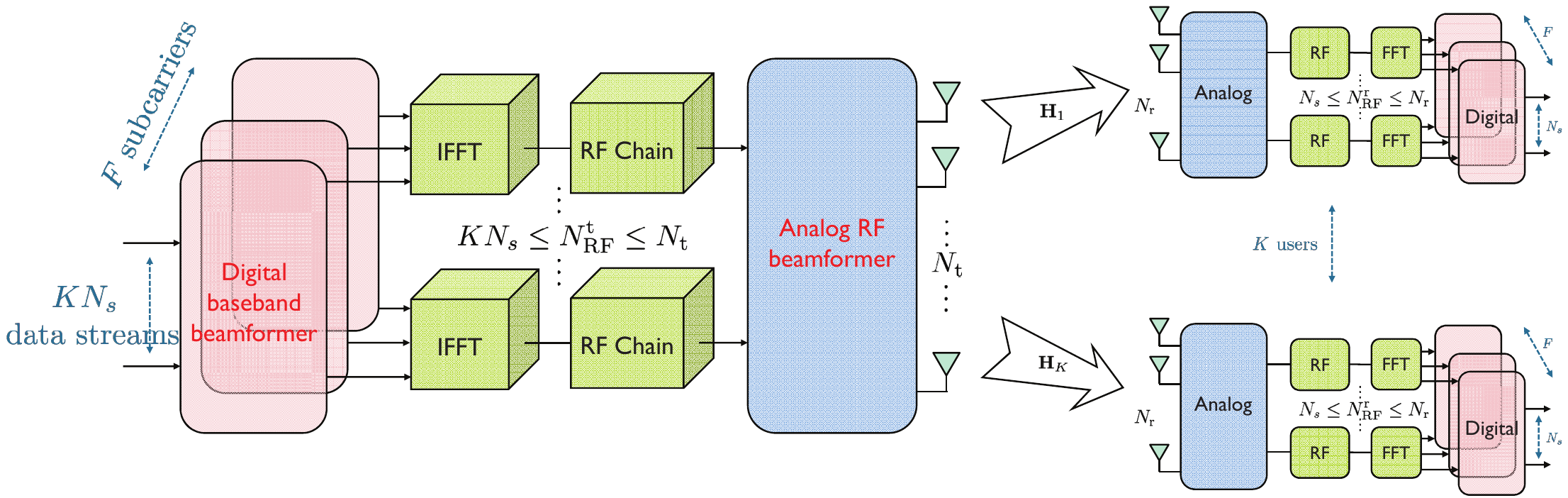}
	\caption{Hybrid beamforming in a general multiuser multicarrier system. The base station (BS), on the left hand side of the figure, is equipped with $\Nt$ antennas to serve $K$ $N_\mathrm{r}$-antenna users over $F$ subcarriers. In addition, $N_s$ data streams are transmitted to each user on each subcarrier. The available numbers of RF chains are $\NRF$ and $N_\mathrm{RF}^\mathrm{r}$ for the BS and each user, respectively.}
	\label{fig5}
\end{figure*}

To deploy mm-wave systems with large-scale antenna arrays, challenges in hardware implementation and algorithm design need to be addressed. A large number of hardware components will be needed to support conventional digital beamforming, including signal mixers, analog-to-digital/digital-to-analog converters (ADCs/DACs), and power amplifiers \cite{6732923}. This will put prohibitive burdens on cost and power consumption, especially for mobile terminals, and thus is not feasible. Furthermore, the significantly increased dimension of the beamformers brings stringent requirements on computational efficiency of beamforming algorithms. These challenges have driven the recent efforts in developing hardware efficient transceivers, supported with efficient beamforming algorithms. One initial proposal is \emph{analog beamforming}, supported with a phase shifter network and low-complexity beam steering. It is currently the de-facto approach for indoor mm-wave systems \cite{6108325}. However, analog beamforming only supports single-stream transmission, and cannot fully exploit the available spatial resource. To further improve performance, \emph{hybrid beamforming} has been proposed as a cost-effective approach to support spatial multiplexing with a limited number of radio frequency (RF) chains, whose potential has been demonstrated in many recent studies \cite{6979963,7387790}. In particular, compared with analog beamforming, hybrid beamforming supports multi-stream transmission with spatial multiplexing, as well as spatial division multiple access. It achieves spectral efficiency comparable to fully digital beamforming with much reduced hardware complexity. Therefore, it has been regarded as a promising candidate for transceiver structures in mm-wave systems.

The concept of hybrid beamforming can be traced back to early 2000s \cite{8030501,1519678}, where the point-to-point single-stream  transmission in sub-6 GHz systems was investigated as a special case. Almost a decade later, hybrid beamforming was revisited in mm-wave systems \cite{6717211} and has drawn considerable attention from both academia and industry. By leveraging the sparsity of mm-wave channels, low-complexity algorithms were proposed for point-to-point hybrid beamforming \cite{6717211}, whose achievable spectral efficiency was further improved in \cite{7417342}. Then, hybrid beamforming was extended to single-user multicarrier \cite{7448873,6884253,7397861} and multiuser single-carrier systems \cite{7335586,7037444,6928432,7913599}.
The main challenge of hybrid beamforming design is to optimize the system performance under the hardware constraints, e.g., reduced RF chains and the high-dimensional phase shifter-based analog beamformer. Various algorithms were developed to combat this difficulty, e.g., compressive sensing \cite{6717211,6955826}, codebook-based design \cite{6874567,6884253}, and manifold optimization \cite{7397861,8616797}, which have offered effective design methodologies for hybrid beamforming.

Nevertheless, hybrid beamforming is still facing several critical issues that may hinder its practical applicability. Compared with fully digital beamforming, hardware complexity has been significantly reduced, but it is still quite a concern, especially considering the cost and power consumption of mm-wave devices \cite{han2015large}. Thus, hybrid beamforming structures that are more hardware-efficient should be developed. For this aspect, we can learn little from conventional digital beamforming design, which takes a performance-oriented perspective, e.g., to maximize spectral efficiency or minimize transmit power, but largely neglects hardware complexity. Furthermore, digital beamforming problems are typically convex, and powerful tools from convex optimization can be leveraged \cite{5447076}. However, hybrid beamforming problems are innately non-convex and challenging to design.

To address these design challenges for hybrid beamforming, a holistic approach should be taken. In particular, we need a comprehensive consideration that accounts for the following three decisive aspects: \emph{hardware efficiency}, \emph{computational efficiency}, and \emph{spectral efficiency}. Accordingly, this paper presents key proposals of hybrid beamforming structures, emphasizing the following three desirable properties:
\begin{enumerate}
	\item	\emph{High hardware efficiency (HE)}, i.e., with as few hardware components as possible, which leads to low cost and low power consumption.
	
	\item	\emph{High spectral efficiency (SE)}, which should be close to that of the fully digital beamforming.
	
	\item	\emph{High computational efficiency (CE)}, i.e., the hybrid beamforming algorithm should be of low complexity.
\end{enumerate}

A special emphasis is placed on the interplay between hardware structures and beamforming algorithm design. Answers to the following key questions will be revealed through the discussion:
\begin{itemize}
\item \emph{How many RF chains and phase shifters are needed?}

\item \emph{Can hybrid beamforming approach the performance of fully digital beamforming?}

\item \emph{How to effectively design hybrid beamforming algorithms?}
\end{itemize}
Specifically, we first present the state-of-the-art hybrid beamforming structures, as well as their algorithm design. Limitations of these basic structures are identified. Then, we introduce two new analog network implementations, which greatly simplify algorithm design and reduce hardware complexity, respectively. Finally, we propose a flexible mapping strategy for hybrid beamforming, which helps to strike a good balance between the hardware complexity and spectral efficiency. The paper ends with key conclusions and some future research directions.

\emph{Notations:}
The following notations are used throughout this paper. $\jmath=\sqrt{-1}$ is the imaginary unit; $\mathbb{C}$ and $\mathbb{Z}$ denote the sets of complex numbers and integer numbers; $\mathbf{a}$ and $\mathbf{A}$ stand for a column vector and a matrix, respectively; The $i$-th row, the $j$-th column, and the $(i,j)$-th entry in matrix $\mathbf{A}$ are denoted as $\mathbf{A}(i,:)$, $\mathbf{A}(:,j)$, and $\mathbf{A}(i,j)$, respectively;
The conjugate, transpose and conjugate transpose of $\mathbf{A}$ are represented by $\mathbf{A}^*$, $\mathbf{A}^T$ and $\mathbf{A}^H$;  $\left\Vert\mathbf{a}\right\Vert_0$ stands for the $\ell_0$-norm of vector $\mathbf{a}$; $\mathrm{blkdiag}(\mathbf{A}_1,\cdots,\mathbf{A}_i)$ establishes a block diagonal matrix using $\mathbf{A}_1,\cdots,\mathbf{A}_i$ as its diagonal terms.

\section{A Primer on Hybrid Beamforming}\label{SecII}
A  hybrid beamforming transceiver is depicted in Fig. \ref{fig5}.
We consider the downlink transmission of a multiuser mm-wave MIMO-OFDM (orthogonal frequency-division multiplexing) system. A base station (BS) leverages an $\Nt$-size antenna array to serve $K$ users over $F$ subcarriers.
The BS transmits $N_s$ data streams to each user on each subcarrier.
The number of available RF chains at the BS is {\color{black}$\NRF$}, which is restricted as $KN_s\le \NRF<\Nt$.\footnote{The settings for hybrid beamforming at the user side can be defined in a similar way as those at the BS side, which are omitted here to keep the presentation clear and concise.}


The hybrid beamformer consists of two components: a digital component and an analog component. The digital part is composed of RF chains, whose structure is common for different proposals to be discussed. Similar to the conventional fully digital beamforming, the digital component in hybrid beamforming can be performed for each user on each subcarrier, denoted as ${\mathbf{F}_\mathrm{BB}}_{k,f}\in\mathbb{C}^{N_\mathrm{RF}^\mathrm{t}\times N_s}$. However, this is not the case for the analog component, or \emph{the analog network}, in hybrid beamforming. Since the transmitted signals for all the users are mixed together by the digital beamformers, and analog RF beamforming is a post-IFFT (inverse fast Fourier transform) operation, the analog network $\FRF\in\mathbb{C}^{\Nt\times N_\mathrm{RF}^\mathrm{t}}$ is a common component shared by all the users and subcarriers.

Furthermore, as will be revealed in this paper, the analog network is the key differentiating compoent in different hybrid beamforming structures. In particular, the structure of the analog network not only influences the hardware efficiency, but also has a significant impact on both the algorithmic design and achievable spectral efficiency. Hence, our discussion mainly focuses on the analog network. In this section, we first introduce key hardware components, and then introduce a new taxonomy for comparing different hybrid beamforming structures.

\subsection{Key Hardware Components}
Hardware efficiency is a key consideration when designing hybrid beamforming structures, and we compare different structures by the number of required key components. Note that, given the rapid advances in hardware and diversified choices, it is difficult to make a fair comparison for energy efficiency, which, nevertheless, will be largely determined by hardware efficiency. Therefore, we do not explicitly consider energy efficiency in this paper.

In the analog RF domain, key hardware components include power amplifiers, phase shifters, and switches. Power amplifiers, as basic components in conventional fully digital beamforming, are needed for each antenna element, and great attention has been drawn on realizing low power amplifiers in integrated circuit (IC) design. In contrast, phase shifters, originally utilized in military radar systems, are the newly-introduced hardware components in hybrid beamforming systems. Hardware suppliers are not yet ready to provide phase shifters for commercial use, and the cost of phase shifters is currently very high, e.g., it can be around a hundred US dollars even with low resolution\footnote{\url{http://www.analog.com/en/parametricsearch/10700\#}}. It motivates alternative structures to replace phase shifters with other components or to reduce the number of phase shifters. For example, Roi \emph{et al.} \cite{7370753} proposed to replace phase shifters with switches to reduce the hardware complexity. Other proposals will be discussed later in the paper.

As power amplifiers are necessary and cannot be easily replaced, the hardware efficiency of the analog network primarily depends on phase shifters and/or switches. As a matter of fact, switches entail only binary states and therefore outperform phase shifters in terms of implementation complexity, power consumption, and cost. However, limiting to the on-off state will inevitably incur performance loss in spectral efficiency. Later we will show how to combine phase shifters and switches to develop hardware-efficient hybrid beamforming structures with good spectral efficiency.

\begin{table*}[t]
	\centering
	\begin{subtable}[h]{1\textwidth}\centering\caption{Mapping strategies for hybrid beamforming.}
		\begin{tabular}{| m{2.55cm} | c | c |c|}\hline
			& \multicolumn{1}{c|}{\textbf{Fully-Connected}} & \multicolumn{1}{c|}{\textbf{Partially-Connected}} & \multicolumn{1}{c|}{\textbf{Group-Connected}} \bigstrut\\\hline
			\textbf{Mapping strategy} &
			\begin{minipage}{.235\textwidth}
				\vspace{.5em}\includegraphics[height=5cm]{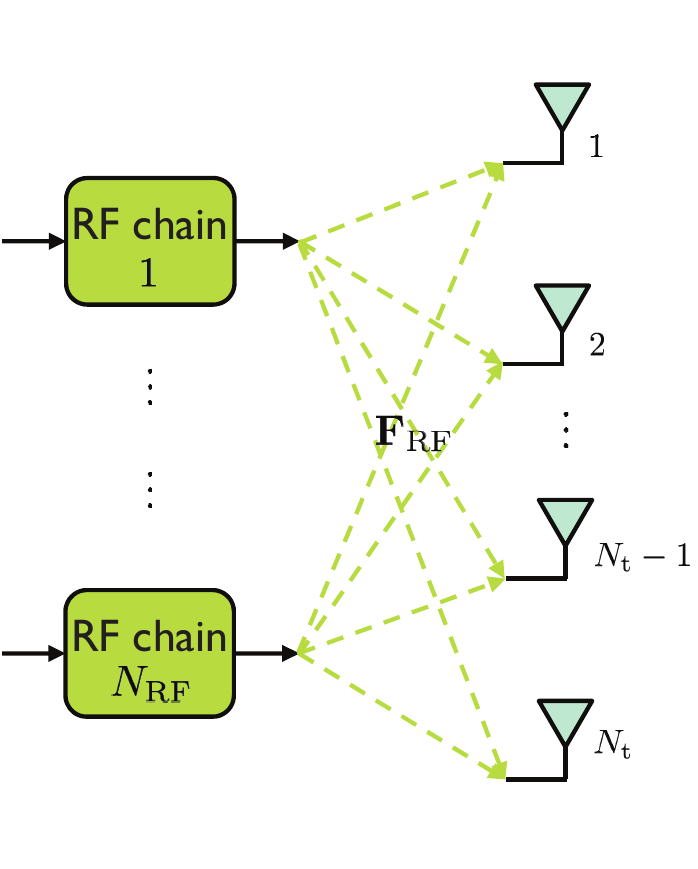}
			\end{minipage}
			&
			\begin{minipage}{.245\textwidth}
				\vspace{.5em}\includegraphics[height=5cm]{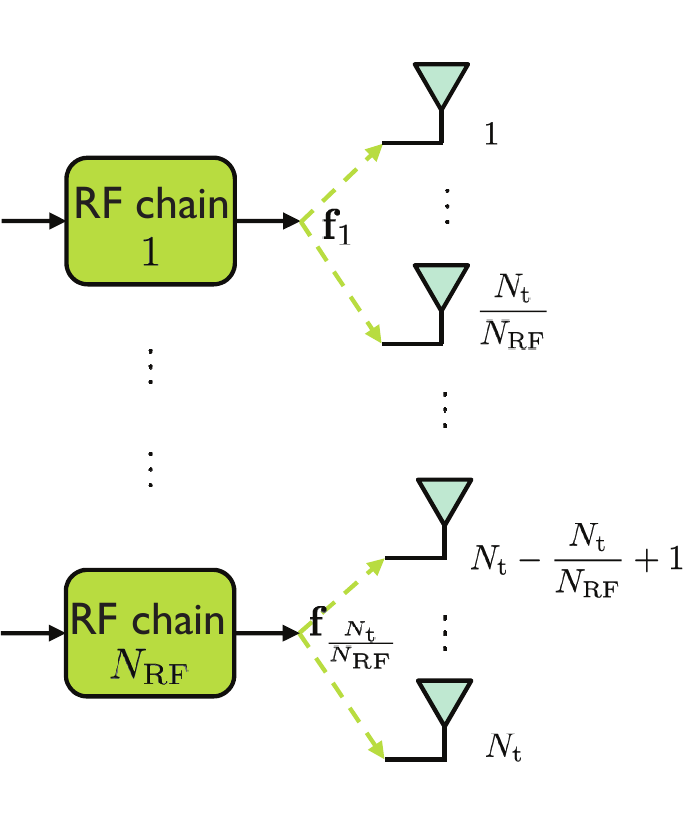}
			\end{minipage}
			&
			\begin{minipage}{.27\textwidth}
				\includegraphics[height=5cm]{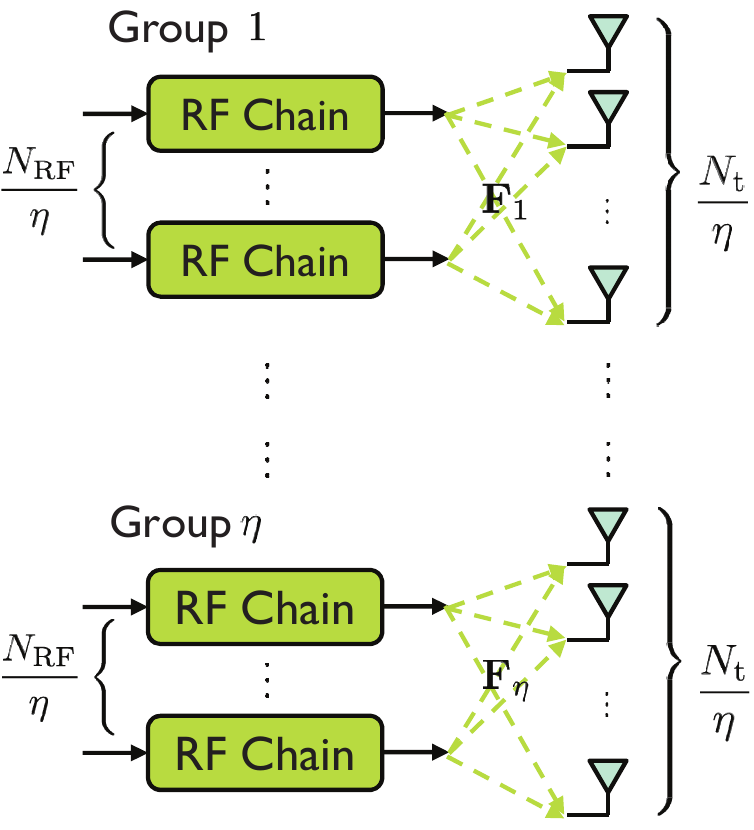}
			\end{minipage}
			\bigstrut\\\hline
			\textbf{Analog beamforming matrix $\FRF$} &
			\begin{minipage}{.235\textwidth}
				\centering{Full matrix}\\
				$\begin{bmatrix}
					f_{11}&f_{12}&\cdots&f_{1\NRF}\\
					f_{21}&f_{22}&&\\
					\vdots&&\ddots&\\
					f_{\Nt1}&&&f_{\Nt\NRF}\\
				\end{bmatrix}$
			\end{minipage}
			&
			\begin{minipage}{.245\textwidth}
				\vspace{0.2em}
				\centering Block diagonal matrix
				$\begin{bmatrix}
				\mathbf{f}_1&&&\\
				&\mathbf{f}_2&&\\
				&&\ddots&\\
				&&&\mathbf{f}_{\frac{\Nt}{\NRF}}
				\end{bmatrix}$
			\end{minipage}
			&
			\begin{minipage}{.27\textwidth}
				\centering Block diagonal matrix
				$\begin{bmatrix}
				\mathbf{F}_1&&&\\
				&\mathbf{F}_2&&\\
				&&\ddots&\\
				&&&\mathbf{F}_\eta
				\end{bmatrix}$
			\end{minipage}
			\bigstrut\\\hline
		\end{tabular}%
	\end{subtable}
	\bigskip
	\begin{subtable}[h]{1\textwidth}\centering\vspace{2em}\caption{Hardware implementations for hybrid beamforming.}
		\begin{tabular}{| m{2.55cm} | c | c |c|}\hline
			& \multicolumn{1}{c|}{\textbf{Single Phase Shifter (SPS)}} & \multicolumn{1}{c|}{\textbf{Double Phase Shifter (DPS)}} & \multicolumn{1}{c|}{\textbf{Fixed Phase Shifter (FPS)}} \bigstrut\\\hline
			\textbf{Hardware\quad  implementation} &
			\begin{minipage}{.23\textwidth}
				\vspace{0.1em}\centering\includegraphics[height=2.5cm]{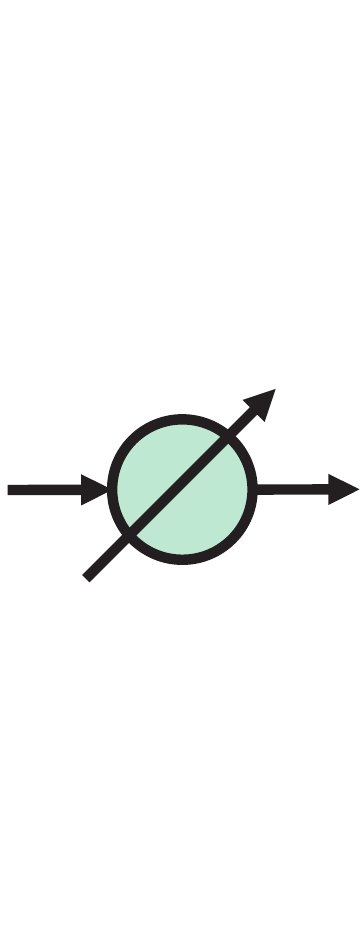}
			\end{minipage}
			&
			\begin{minipage}{.23\textwidth}
				\vspace{0.1em}\centering\includegraphics[height=2.5cm]{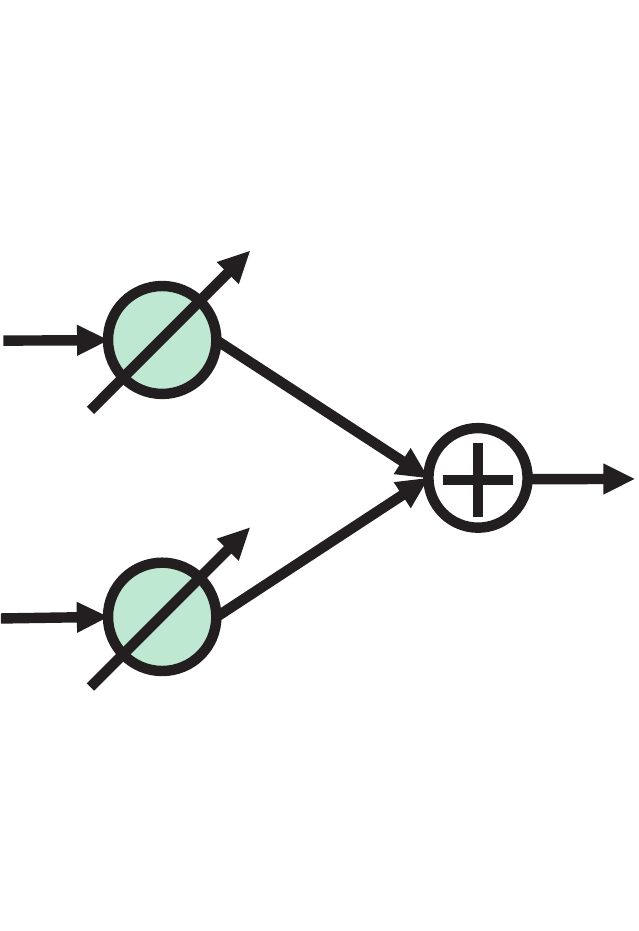}
			\end{minipage}
			&
			\begin{minipage}{.23\textwidth}
				\vspace{0.2em}\centering\includegraphics[height=2.5cm]{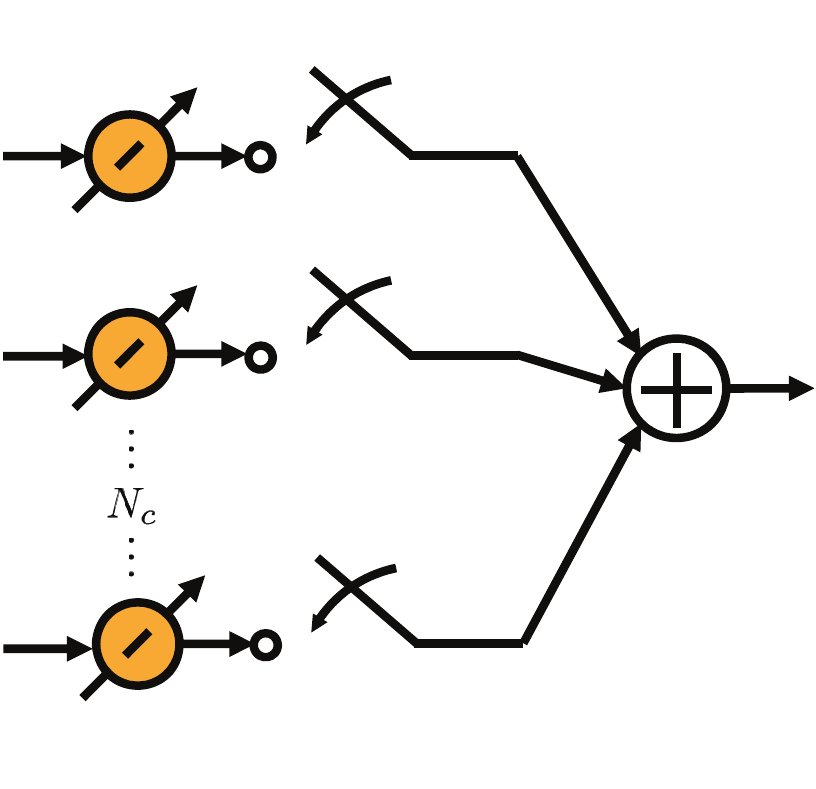}
			\end{minipage}
			\bigstrut\\\hline
			\multirow{2}[0]{*}{\textbf{Constraint on  $\FRF$}} & \multirow{2}[0]{*}{$|\FRF(i,j)|=1$} & \multirow{2}[0]{*}{$|\FRF(i,j)|\le2$} & $\FRF(i,j)=\sum_{n=1}^{N_c}b_ne^{j\theta_n},$ \\
			&       &       &  $b_n\in\{0,1\}$\bigstrut\\\hline
			\multirow{2}[0]{*}{\textbf{Comments}} & One phase shifter for each
			&  Two phase shifters for each		&  $N_c$ fixed phase shifters shared\\
			&  RF chain-antenna pair      & RF chain-antenna pair      &  by all RF chain-antenna pairs\bigstrut\\\hline
		\end{tabular}%
	\end{subtable}
	\caption{\emph{Mapping strategies} and \emph{hardware implementations} for hybrid beamforming, with $\FRF$ as the analog beamforming matrix, and $\mathbf{f}_i$ and $\mathbf{F}_i$ denoting a column vector and a matrix, respectively. To realize a specific analog network structure, one may pick a mapping strategy from (a) to decide how to connect the RF chain-antenna pairs. Then, one should choose a hardware implementation from (b) to realize each RF chain-antenna pair.}
\end{table*}

\subsection{A Taxonomy of Hybrid Beamforming Structures}
Hybrid beamforming structures differ mainly in the way they use the above-mentioned hardware components to compose the analog network.
In particular, the analog network structure is primarily determined by two elements, i.e., the \emph{mapping strategy} and \emph{hardware implementation}, for which different proposals are listed in Tables I (a) and (b).
\begin{itemize}
\item \textbf{The mapping strategy}: It determines how the RF chains and antenna elements are connected. As shown in Table I (a), there are two basic mapping strategies, namely, the fully- and partially-connected mapping, which will be introduced in Section \ref{Sec:basic structure}. A more flexible mapping strategy, group-connected mapping, will be introduced in Section \ref{SecIV}. Table I (a) further shows the analog beamforming matrix associated with each mapping strategy, which bears a special structure that will affect the beamformer design.
\item \textbf{The hardware implementation}: It specifies the adopted hardware components and the way each RF chain-antenna pair is connected. Among the three implementations shown in Table I (b), the single phase shifter (SPS) implementation is the most commonly used one, and the other two, double phase shifter (DPS) and fixed phase shifter (FPS) implementations, are recently proposed and will be introduced in later sections. Different hardware implementations will induce different constraints on $\FRF$, as shown in the table, which will significantly affect the algorithm design.
\end{itemize}
As a common example, the SPS fully-connected structure refers to the fully-connected mapping strategy with a single phase shifter connecting each RF chain with a corresponding antenna.

\section{Basic Hybrid Beamforming Structures}\label{Sec:basic structure}
In this section, we present and compare two basic mapping strategies, namely, the \emph{fully-} and \emph{partially-connected} ones.
As shown in Table I (a), in the fully-connected mapping strategy, every RF chain is connected to all the antenna elements, while each RF chain is connected to a subset of neighboring antenna elements that do not overlap with each other  in the partially-connected mapping strategy.
For hardware implementation, we consider the classic SPS implementation, i.e., each connected RF chain-antenna pair is linked via a single phase shifter. Therefore, in terms of hardware efficiency, the SPS fully- and partially-connected ones employ $\Nt\NRF$ and $\Nt$ phase shifters, respectively. Through the following comparison of these two basic structures, we shall illustrate their limitations and motivate other proposals.

\subsection{Basic Principles of Hybrid Beamforming Algorithm Design}
In this part, we present a basic formulation for hybrid beamforming design, accompanied with some design principles.
A common design principle is to approximate the fully digital beamformer subject to the constraint for the analog beamforming matrix  \cite{6717211,7397861}, whose formulation is correspondingly given by
\begin{equation}\label{problemformulation}
\begin{aligned}
&\underset{\mathbf{F}_\mathrm{RF},\mathbf{F}_\mathrm{BB}}{\mathrm{minimize}} && \left\Vert \mathbf{F}_\mathrm{opt}-\mathbf{F}_\mathrm{RF}\mathbf{F}_\mathrm{BB}\right\Vert _F\\
&\mathrm{subject\thinspace to}&&
\begin{cases}
\left\|\mathbf{F}_\mathrm{RF}\mathbf{F}_\mathrm{BB}\right\|_F^2\le KN_sF\\
\FRF\in\mathcal{A},
\end{cases}
\end{aligned}
\end{equation}
where the combined fully digital beamformer is denoted as $\Fopt=\left[{\Fopt}_{1,1},\cdots,{\Fopt}_{k,f},\cdots,{\Fopt}_{K,F}\right]\in\mathbb{C}^{\Nt\times KN_sF}$, and $\FBB=\left[{\FBB}_{1,1},\cdots,{\FBB}_{k,f},\cdots,{\FBB}_{K,F}\right]$ is the concatenated digital beamformer with dimension ${\NRF\times KN_sF}$.
The first constraint in the formulation is the total transmit power constraint, and the second constraint depends on the adopted hardware implementation, as shown in Table I. The main merits of this formulation include its general applicability, i.e., it can be applied with any given digital beamformer, and the tractability for algorithm design, to be illustrated below.

\emph{\textbf{Critical Role of the Analog Network}}: In the hybrid beamforming design problem \eqref{problemformulation}, $\mathcal{A}$ is the feasible set of the analog network, which is distinct for different hybrid beamforming structures.
Before we proceed, we would like to emphasize the critical role of the analog network. As we discussed before, the analog network is shared by all the users and subcarriers, so a single analog beamforming matrix should match the channel states of different users on different subcarriers. This is an extremely difficult task, and it is not clear at all how close we can approach the performance of the fully digital beamforming with hybrid beamforming. With such a decisive role on achievable performance, the analog network calls for a delicate design. Moreover, different implementations of the analog network bring different constraints for the analog beamforming matrix, and thus determine the difficulty in beamforming algorithm design. Both of these aspects will be elaborated throughout the discussion in this paper.

 As there are two components in a hybrid beamformer, i.e., an analog one and a digital one, \emph{alternating minimization} (AltMin) serves as a basic design principle \cite{7397861}. It alternately optimizes the analog and digital parts. It is apparent that the optimization of the digital beamforming matrix $\FBB$ is a least squares problem, which has a closed-form solution. On the other hand, with the SPS implementation, the main difficulty lies in the analog component, for which there is a non-convex unit modulus constraint. In particular,  the feasible set $\mathcal{A}$ of the analog network $\FRF$  can be specified by a set of matrices where the amplitude of each non-zero element is forced to be 1, i.e., $\left|\FRF({i,j})\right|=1$ \cite{6717211}. Design methodologies for the two basic structures are different, as presented in the following two subsections.

\subsection{SPS Fully-Connected Structure}
Note that when $\NRF\ge 2KN_s$, the fully digital beamforming can be realized by the SPS fully-connected structure \cite{1519678,zhang2014achieving}, and this case is trivial in terms of algorithm design. For the general case when $\NRF<2KN_s$, the orthogonal matching pursuit (OMP) algorithm \cite{6717211} is the most widely-used algorithm, which treats the analog network design as a sparsity constrained matrix reconstruction problem. In particular, the columns of the analog beamforming matrix $\FRF$ are selected from a candidate set, which typically consists of the array response vectors of mm-wave channels. This codebook-based design inevitably incurs some performance loss when approaching the fully digital beamforming.
More recent attention focused on reducing the computational complexity of the OMP algorithm, e.g., by reusing the matrix inversion result in each iteration \cite{6955826}.

In \cite{7397861}, by recognizing that the unit modulus constraints of the analog network define a complex circle Riemannian manifold, a  manifold optimization based AltMin (MO-AltMin) algorithm was proposed, which outperforms the OMP algorithm but with increased complexity.
In particular, by defining key elements, e.g., inner products and gradients, in the neighborhood of a manifold that is homeomorphic to the Euclidean space, a variety of classic optimization algorithms in the Euclidean space can be transplanted to manifold optimization. For instance, the conjugate gradient method in the Euclidean space was adopted on the complex circle manifold for hybrid beamforming in \cite{7397861,7417342}.

As introduced above, the OMP algorithm updates a column of the analog beamforming matrix $\FRF$ at a time while the MO-AltMin algorithm optimizes the whole $\FRF$ matrix in each iteration. To the other extreme, the phase shifters are optimized one by one in \cite{7389996}. In  particular, the contribution of each phase shifter to the spectral efficiency was analytically identified, based on which the analog network was iteratively optimized in a phase shifter-by-phase shifter fashion. This approach also suffers a high complexity since the number of iterations of the algorithm is proportional to the number of phase shifters in use, which is typically a huge number ($\Nt\NRF$) in mm-wave MIMO systems with the SPS fully-connected structure.

\begin{table*}[t]
	
	{\color{black}
		\caption{Existing works with the SPS implementation}
	\begin{center}
		\begin{tabular}{|c|c|c|c|c|}
			\hline
			\multicolumn{1}{|c|}{\textbf{Hybrid beamforming}}                       & \textbf{Number of    }     & \multicolumn{1}{c|}{\textbf{Design}}  & \multicolumn{1}{c|}{\textbf{Spectral}}      & \multirow{2}{*}{\textbf{Computational complexity}}  \bigstrut\\ 
			\textbf{structure}                                                      & \textbf{phase shifters}    &     \textbf{Approach}                                &   \textbf{efficiency}                                   &                                                                \bigstrut \\ \hhline{|=|=|=|=|=|}
			\multicolumn{1}{|c|}{\multirow{3}{*}{SPS fully-connected}}                    &     \multirow{3}{*}{$\Nt\NRF$            }               & OMP       \cite{6717211}                          &        \checkmark\checkmark\checkmark                              &    $\mathcal{O}\left(\Nt\NRF\left(L N_s+{\NRF}^2+2\NRF N_s\right)\right)^\dagger$                                                           \bigstrut \\ \cline{3-5} 
			\multicolumn{1}{|c|}{}                                         &                  & MO-AltMin  \cite{7397861}                         &         \checkmark\checkmark\checkmark\checkmark\checkmark                             &  Extremely high                                                       \bigstrut \\ \cline{3-5} 
			\multicolumn{1}{|c|}{}                                         &       & Element-wise   \cite{7389996}                 &    \checkmark\checkmark\checkmark\checkmark                               &     $\mathcal{O}\left(N_\mathrm{iter}\Nt^4\NRF\right)$                                                                  \bigstrut\\ \hline
			\multicolumn{1}{|c|}{\multirow{2}{*}{SPS partially-connected}} & \multirow{2}{*}{$\Nt$   } & SIC-based    \cite{7445130}                       &      \checkmark                                &     $\mathcal{O}\left(\frac{\Nt^2}{{\NRF}^2}\left(\NRF N_\mathrm{iter}+N_\mathrm{r}\right)+2\NRF N_\mathrm{iter}\right)$                                                            \bigstrut \\ \cline{3-5} 
			\multicolumn{1}{|c|}{}                                         &                   & SDR-AltMin     \cite{7397861}                     &      \checkmark\checkmark                                &                $\mathcal{O}\left(N_\mathrm{iter}{\NRF}^3N_s^3\right)$                                                \bigstrut  \\ \hline
		\end{tabular}\label{tableII}
	\end{center}
\begin{tablenotes}
	\item ${}^\dagger$ $L$ is the total number of paths in the channels.
	 \item * More \checkmark means higher spectral efficiency and $N_\mathrm{iter}$ denotes the number of iterations involved in the algorithm. The computational complexity is evaluated for single-user single-carrier systems for fair comparison. As there are nested iterations in the MO-AltMin algorithm, its computational complexity is much higher than those of other algorithms. 
\end{tablenotes}
}
\end{table*}

\subsection{SPS Partially-Connected Structure}
While most initial efforts on hybrid beamformer design were on the SPS fully-connected structure, the SPS partially-connected one has attracted more recent attention due to its low hardware complexity. In the analog RF domain, the hardware complexity of the SPS partially-connected structure is the same as that of analog beamforming, as the numbers of phase shifters are both equal to the antenna size.
In \cite{7037400,6824962}, codebook-based design of hybrid beamformers was presented for narrowband and OFDM systems, respectively.
Although the codebook-based design enjoys a low complexity, there will be certain performance loss, and it is not clear how much performance gain can be further obtained.
Another proposal is based on the concept of successive interference cancellation (SIC) \cite{7445130}. It decomposes the total achievable rate optimization problem into a series of simple sub-rate optimization problems, each of which only considers the antenna elements connected to one RF chain. However, this approach enforces that the digital beamforming matrix is diagonal, and the number of RF chains should be equal to that of the data streams.

More recently, a semidefinite relaxation based AltMin (SDR-AltMin) algorithm was proposed in \cite{7397861}. This algorithm effectively designs the hybrid beamformer by offering globally optimal solutions for both subproblems of analog and digital beamformers in each alternating iteration, and thus achieves very good performance. In particular, the hybrid beamformer design problem is decoupled for each RF chain and its connected antenna elements. In this way, each subproblem is reformulated as a non-convex quadratically constrained quadratic programming (QCQP) problem, to which the SDR approach was applied, and the tightness of such an SDR is proved in \cite{7397861}.
\begin{figure*}[t]
	\centering
	\begin{minipage}[t]{.49\textwidth}
		\centering\includegraphics[height=6.5cm]{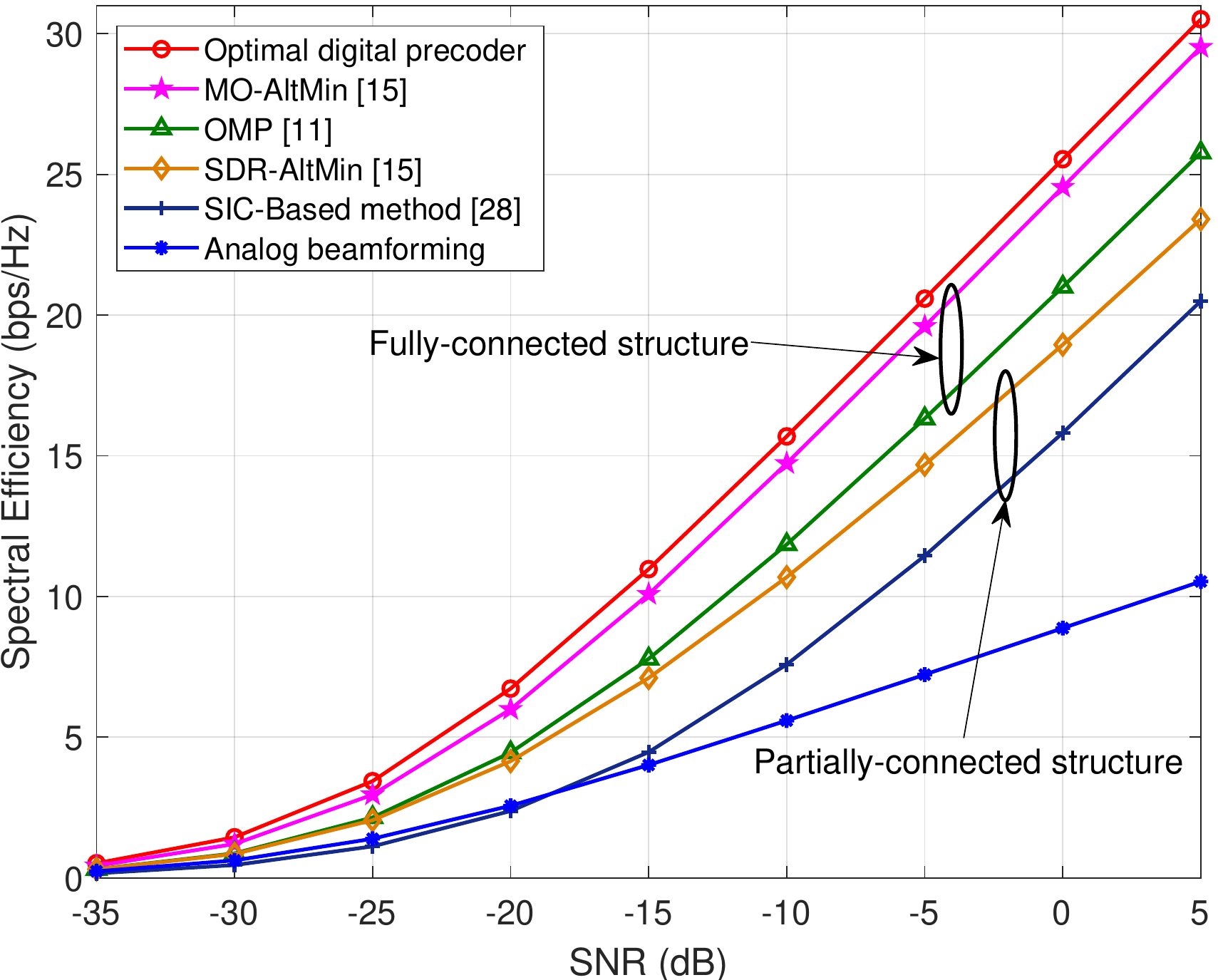}
		\caption{{\color{black}Spectral efficiency of different beamforming algorithms for a point-to-point link in a flat-fading channel. The BS and user are equipped with 144 and 36  antennas, respectively, to transmit 3 data streams with 3 RF chains at both the transmitter and receiver sides.}}
		\label{fig0}
	\end{minipage}\hfill
	\begin{minipage}[t]{.49\textwidth}
		\centering\includegraphics[height=6.5cm]{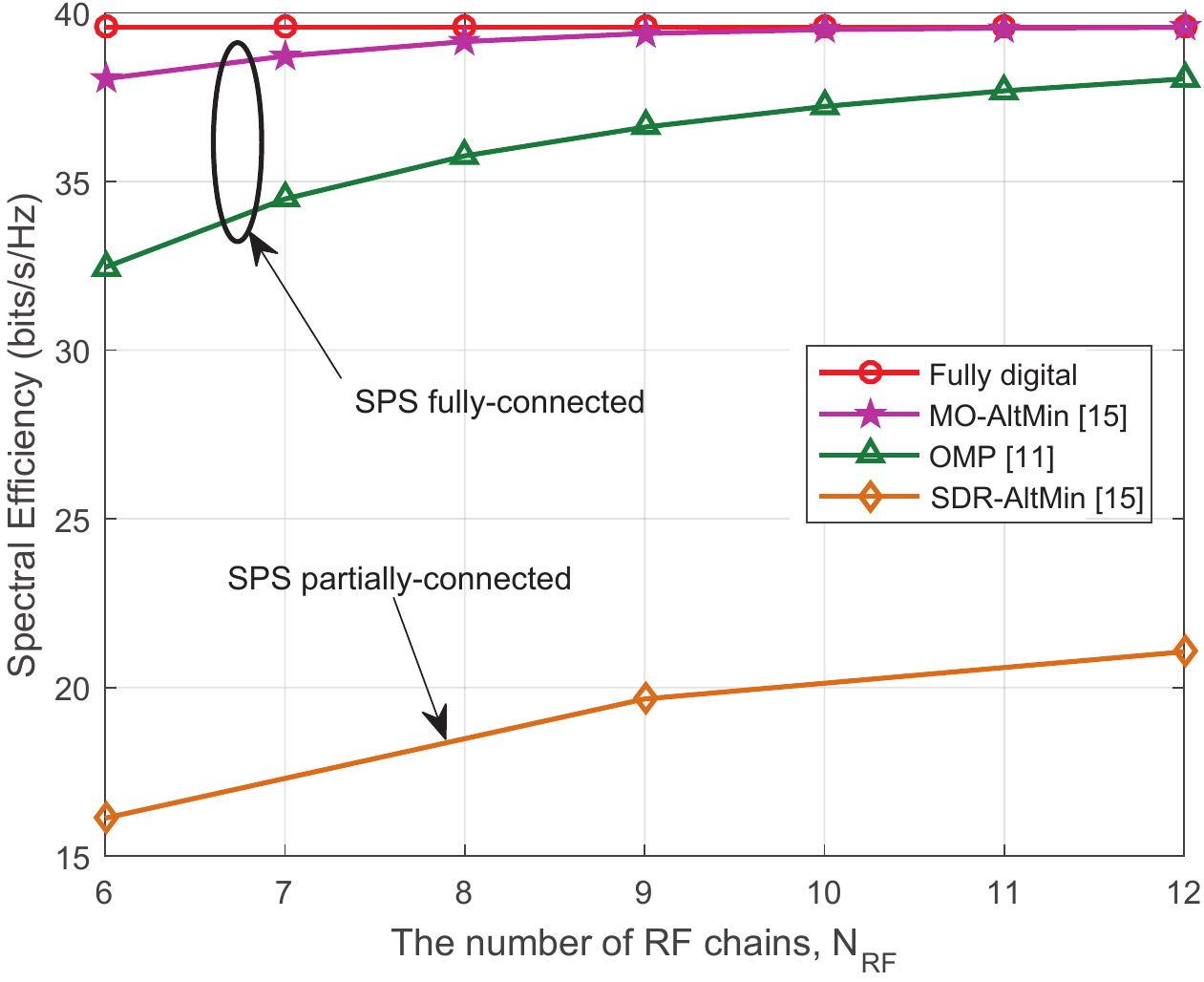}
		\caption{Spectral efficiency of different beamforming algorithms for a point-to-point link in a flat-fading channel. The BS and user are equipped with 144 and 36  antennas, respectively, to transmit 6 data streams with SNR as 0 dB. The x-axis stands for the numbers of RF chains at both the transmitter and receiver sides.}
		\label{fig1}
	\end{minipage}
\end{figure*}

{\color{black}
	The achievable spectral efficiency of existing representative hybrid beamforming algorithms is compared in Fig. \ref{fig0}. As can be observed, the MO-AltMin algorithm achieves the highest spectral efficiency for the SPS  fully-connected structure while the SDR-AltMin algorithm outperforms other benchmarks for the SPS partially-connected structure.
	Furthermore, the numbers of hardware components and computational complexity of corresponding design algorithms for hybrid beamforming with the SPS implementation are summarized in Table \ref{tableII}. In particular, the partially-connected mapping strategy entails a lower computational complexity thanks to its simpler hardware implementation.
}

\subsection{Limitations of Basic Structures}


We compare spectral efficiency of the two basic structures in Fig. \ref{fig1}. It shows a clear performance gap between the two structures, with the fully-connected structure providing much higher spectral efficiency than the partially-connected one. 
Furthermore, the comparison between the MO-AltMin and OMP algorithms demonstrates the importance of efficient algorithms to reach realistic conclusions. In particular, with the MO-AltMin algorithm, the fully-connected structure is shown to approach the performance of the fully digital one with the number of RF chains comparable to the number of data streams, while the OMP algorithm fails to achieve this. These observations demonstrate that the limited number of RF chains in hybrid beamforming is not a performance bottleneck, but the analog network structure has a decisive effect.

The above comparison reveals several key limitations of the two basic structures.
\begin{itemize}
	\item \textbf{Algorithmic perspective}: While the SPS fully-connected structure with the MO-AltMin algorithm approaches the performance of the fully digital beamforming, its computational complexity is extremely high \cite{7397861}. It is not clear how close we can approach fully digital beamforming with more practical algorithms for this structure.
	\item \textbf{Hardware perspective}: The SPS fully-connected structure has the potential to perform closely to the fully digital one, but still with high hardware complexity in the analog network. The SPS partially-connected structure significantly reduces the number of phase shifters, but with much degraded performance.
\end{itemize}

Therefore, key innovations in both the hardware and algorithmic aspects are needed before we see the commercial success of hybrid beamforming. From the above discussion, we have already observed that the analog network structure greatly affects the algorithm design. So the key challenge is to design the analog network to reduce hardware complexity, as well as enabling low-complexity beamforming algorithms, which will be addressed in Section \ref{SecIII}.

Inevitably, trade-offs need to be made among hardware efficiency, computational efficiency, and spectral efficiency. The two basic structures provide such a trade-off, but in an extreme way. The fully-connected mapping strategy is with too high hardware complexity, as well as algorithm complexity if with the MO-AltMin algorithm, while the partially-connected mapping strategy incurs too much performance degradation. It is thus of practical importance to develop new structures that can achieve more flexible trade-offs. To address this aspect, a flexible mapping strategy will be presented in Section \ref{SecIV}.

\section{Two New Analog Network Implementations}\label{SecIII}
In this section, we introduce two recent proposals for the analog network implementation, which improve upon the SPS implementation in different aspects. The first proposal, namely the \emph{double phase shifter} (DPS) implementation, simplifies the algorithm design and improves spectral efficiency, at the cost of more phase shifters. One byproduct of the investigation of this implementation is a convex relaxation approach to develop highly efficient beamforming algorithms. The second proposal, called the \emph{fixed phase shifter} (FPS) implementation, only requires a small number of fixed phase shifters, supplemented with switches, and thus it improves hardware efficiency. As will be shown later, it also does well in computational efficiency and spectral efficiency.
\subsection{Double Phase Shifter (DPS) Implementation}\label{SecIII-A}
\begin{figure}[t]
	\centering\includegraphics[height=4.8cm]{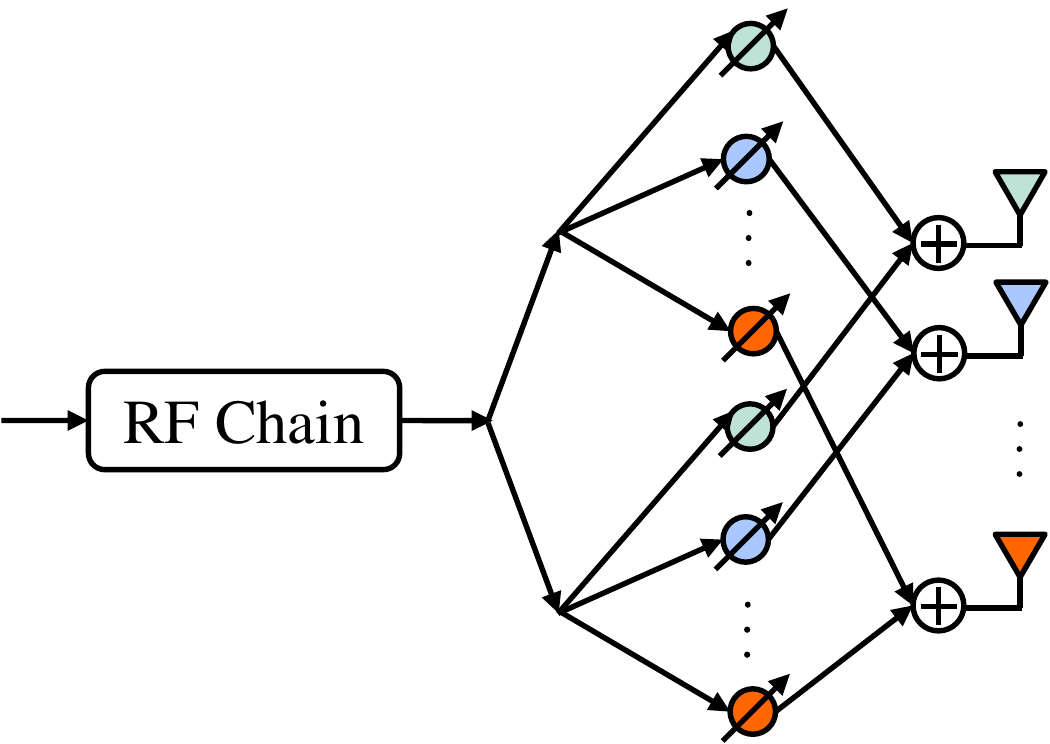}
	\caption{DPS implementation for hybrid beamforming.}
	\label{fig12}
\end{figure}
In this part, we present a new hardware implementation to enable efficient hybrid beamforming algorithms. For the SPS implementation, the unit modulus constraint for the analog network forms the main challenge for algorithm design. The principal obstacle is that we can only adjust the phase but not the amplitude of the RF signals.

To overcome this constraint, the DPS implementation employs two sets of phase shifters, as shown in Fig. \ref{fig12}. Thus, there are $2\Nt\NRF$ and $2\Nt$ phase shifters for the DPS fully- and partially-connected structures, respectively. For each connection from an RF chain to one of its connected antenna elements, one unique phase shifter in each group will be selected and summed up to compose the analog beamforming gain. In this way, each non-zero element in the analog network corresponds to a sum of the outputs of two phase shifters. Correspondingly, the feasible set $\mathcal{A}$ in \eqref{problemformulation} is specified by a set of matrices where the non-zeros entries have amplitudes less than 2, i.e., $|\FRF(i,j)|=|e^{\jmath\phi}+e^{\jmath\theta}|\le 2$, where $\phi$ and $\theta$ are two phase shifts from each group, respectively. Thus, the new constraints of the analog beamforming matrix become convex, which makes beamforming algorithm design more tractable.
This new implementation fundamentally changes the algorithm design, and computationally efficient beamforming algorithms have been developed for both the fully- and partially-connected mapping strategies \cite{doubling}.

\subsubsection{Fully-Connected Mapping}
For the fully-connected mapping, the hybrid beamforming problem can be specified as
\begin{equation}\label{eq2}
\begin{aligned}
&\underset{\mathbf{F}_\mathrm{RF},\mathbf{F}_\mathrm{BB}}{\mathrm{minimize}} && \left\Vert \mathbf{F}_\mathrm{opt}-\mathbf{F}_\mathrm{RF}\mathbf{F}_\mathrm{BB}\right\Vert _F\\
&\mathrm{subject\thinspace to}&&
\begin{cases}
|\FRF(i,j)|\le2\\
\left\|\mathbf{F}_\mathrm{RF}\mathbf{F}_\mathrm{BB}\right\|_F^2\le KN_sF,
\end{cases}
\end{aligned}
\end{equation}
It is proved in \cite{doubling} that the two constraints in \eqref{eq2} are  redundant, and the remaining problem turns out to be a low-rank matrix approximation problem, which has been well studied and is with a closed-form solution.

It has been investigated that the fully digital beamforming can be achieved when $\NRF\ge 2KN_s$ with the SPS fully-connected structure \cite{zhang2014achieving}. In other words, $2KN_s$ RF chains and $2KN_s\Nt$ phase shifters are enough for achieving fully digital beamforming in single-carrier systems. In contrast, the formulation \eqref{eq2} of the DPS fully-connected structure reveals its optimality in single-carrier systems.

\begin{lemma}
	For single-carrier systems, with the DPS implementation, a fully digital beamformer $\Fopt$ can be perfectly decomposed into  $\FRF$ and $\FBB$ using the minimum number of RF chains, i.e., $\NRF=KN_s$ and $\NRF=N_s$.
\end{lemma}
\begin{IEEEproof}
	The proof can be easily obtained by the rank sufficiency of $\FRF$ and $\FBB$ in the decomposition when $F=1$.
	\end{IEEEproof}

This lemma means that $KN_s$ RF chains and $2KN_s\Nt$ phase shifters are enough for achieving fully digital beamforming, which reduces the required number of RF chains by half compared to the state-of-the-art with the SPS implementation. This phenomenon clearly demonstrates the superiority of doubling the phase shifters in the analog network for hybrid beamforming.

{
	\color{black}
	When it comes to multiuser multicarrier systems, typically $KN_sF\ge \Nt$, the rank of $\Fopt$ should be $\Nt$ (instead of $KN_s$ as single-carrier systems)\footnote{Without loss of generality, we assume all the beamforming matrices in \eqref{eq2} have full rank.} and thus perfect decomposition can only be achieved when $\NRF\ge \Nt$, which, however, severely deviates from the setting of hybrid beamforming. In this way, the matrix decomposition cannot be perfect for hybrid beamformer design due to the rank deficiency, i.e., $\NRF=\mathrm{rank}\left(\FRF\FBB\right)\ll\mathrm{rank}\left(\Fopt\right)=\Nt$. Therefore, problem \eqref{eq2} is typically a low-rank matrix approximation problem, with a closed-form solution as
	\begin{equation}\label{eq16}
	\left(\FRF\FBB\right)^\star\triangleq\mathbf{\hat F}_\mathrm{opt}=\mathbf{U}_1\mathbf{S}_1\mathbf{V}_1^H.
	\end{equation}
	Denote the SVD of $\Fopt$ as $\Fopt=\mathbf{USV}^H$, where matrices $\mathbf{U}_1$ and $\mathbf{V}_1$ are the first $\NRF$ columns of $\mathbf{U}$ and $\mathbf{V}$, respectively, and $\mathbf{S}_1$ is the diagonal matrix whose diagonal elements are the $\NRF$ largest singular values of $\Fopt$. This means that the optimal solution of $\FRF\FBB$ is simply obtained by extracting the $\NRF$ most principle components of $\Fopt$. 
}

\textbf{Convex relaxation for efficient hybrid beamforming:} In addition, inspired by the beamformer design of the DPS fully-connected structure,  a convex relaxation approach for the hybrid beamformer design with the SPS fully-connected structure has been developed \cite{doubling}. Assume that the optimal solution to the low-rank approximation problem \eqref{eq2} is  $\mathbf{\hat{F}_\mathrm{opt}}$, and we propose to extract the phases of the optimal analog network for the DPS implementation to construct the SPS solution, given by
\begin{equation}
\FRF=\exp\{\jmath\angle\left(\mathbf{U}_1\right)\}, \quad\FBB=\mathbf{S}_1\mathbf{V}_1^H.
\end{equation}
where $\angle$ extracts the angle information of a complex matrix in an element wise.
Note that the unitary matrix $\mathbf{U}_1$ fully extracts the information of the column space of $\mathbf{\hat{F}_\mathrm{opt}}$, whose basis are the orthonormal columns in $\FRF$.
This approach only requires an singular value decomposition (SVD) operation, which leads to a low-complexity beamforming algorithm by extracting phases from the DPS solution.

\begin{figure}[t]
	\centering\includegraphics[height=6.5cm]{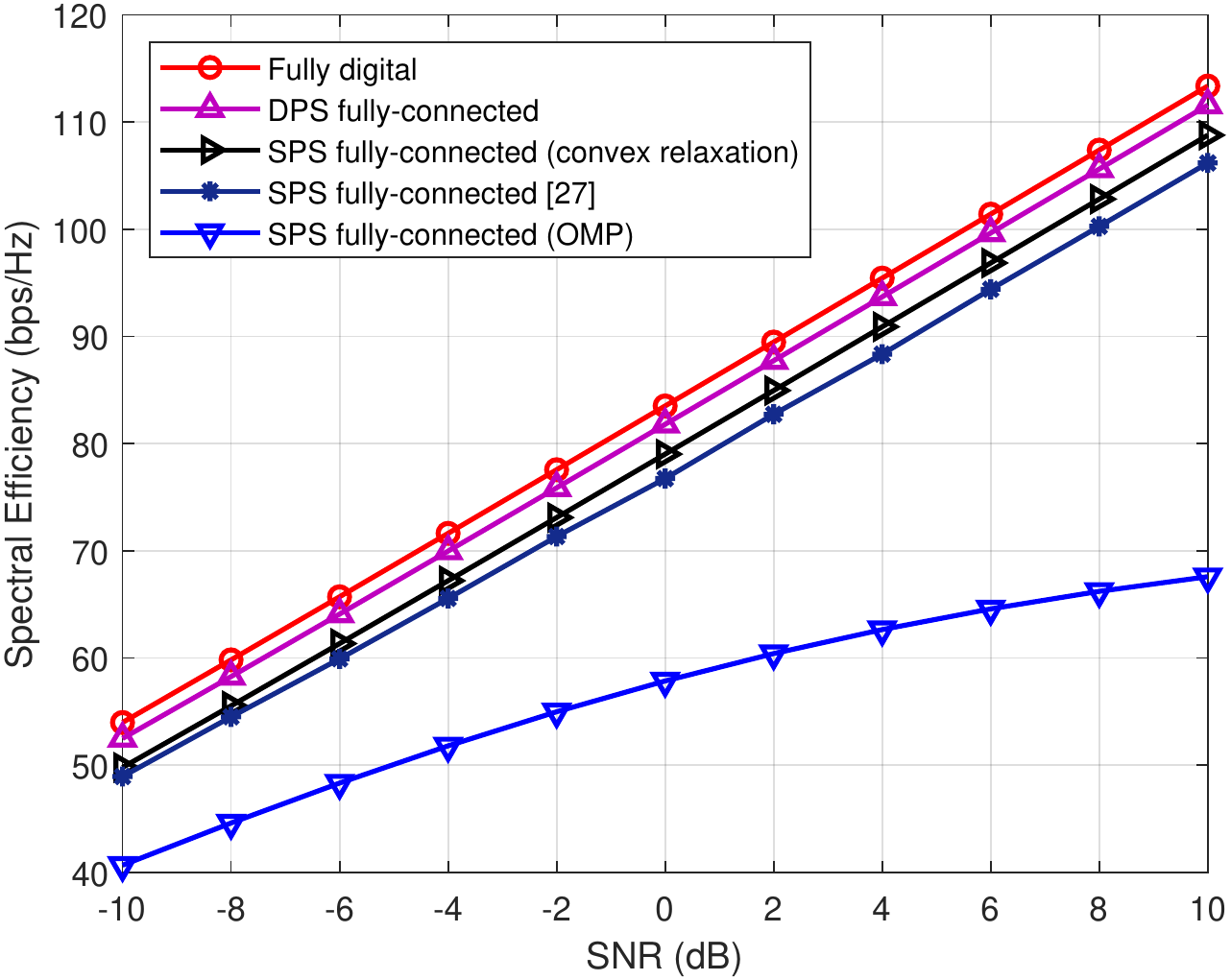}
	\caption{Spectral efficiency achieved by different SPS and DPS fully-connected structures. There are 3 mobile users and the BS transmits 3 data streams to each user. The BS and user are equipped with 256 and 16 antennas, respectively.}
	\label{fignew}
\end{figure}

Fig. \ref{fignew} shows the spectral efficiency achieved by the DPS fully-connected structure, and that of the SPS fully-connected structure with different algorithms.
It shows that the DPS implementation outperforms the SPS implementation, and can achieve a near-optimal performance in terms of spectral efficiency, thanks to the doubling of the phase shifters. In addition, the SPS implementation with the convex relaxation algorithm outperforms the state-of-the-art algorithm in \cite{7389996}, while enjoying much lower computational complexity, which demonstrates the effectiveness of the proposed approach.

\subsubsection{Partially-Connected Mapping}
On the other hand, similar to the SPS partially-connected structure, the hybrid beamforming design with the DPS partially-connected mapping can be decoupled in an RF chain-by-RF chain sense.
The optimization of the hybrid beamformer for the $j$-th RF chain is given by
\begin{equation}
\mathcal{P}_j:\quad\underset{\{a_i\},\mathbf{x}_j}{\mathrm{minimize}}  \sum_{i\in\mathcal{F}_j}
\left\Vert\mathbf{y}_i-a_i\mathbf{x}_j\right\Vert_2^2,
\end{equation}
where $a_i$ is the non-zero element in $\FRF(i,:)$\footnote{{\color{black}Since each antenna element is connected to one RF chain in the partially-connected mapping strategy, there is only one non-zero element in each row of $\FRF$.}}, $\mathcal{F}_j=\left\{i\in\mathbb{Z}\left|{(j-1)\frac{\Nt}{\NRF}+1\le i\le j\frac{\Nt}{\NRF}}\right.\right\}$, $\mathbf{y}_i=\mathbf{F}_\mathrm{opt}^T(i,:)$, and $\mathbf{x}_j=\mathbf{F}_\mathrm{BB}^T(j,:)$.
It is shown in \cite{doubling} that $\mathcal{P}_j$ is an eigenvalue problem. Thus, the DPS implementation brings great advantages in computational efficiency with the closed-form solutions.

The DPS partially-connected structure employs $2\Nt$ phase shifters, which falls in between the numbers of phase shifters in use for the SPS partially-connected structure ($\Nt$) and the DPS fully-connected one ($\NRF\Nt$). To further boost the spectral efficiency with $2\Nt$ phase shifters, a dynamic mapping for the DPS partially-connected structure was proposed in \cite{doubling}. In particular, each RF chain is still connected to a subset of antenna elements, but not necessarily the neighboring ones. In other words, each RF chain is able to select which antenna elements to connect in order to increase the spectral efficiency. For dynamic mapping, the feasible set $\mathcal{A}$ in \eqref{problemformulation} can be specified as  a set of matrices for which every row only has one non-zero entry, i.e., {$\mathcal{A}=\left\{\mathbf{A}|||\mathbf{A}(i,:)||_0=1\right\}$}, and the dynamic mapping design problem is formulated as \cite{doubling}
\begin{equation}\label{eq31}
\begin{aligned}
&\underset{\{\mathcal{D}_j\}_{j=1}^{\NRF}}{\mathrm{maximize}} && \sum_{j=1}^{\NRF}\lambda_1\left(\sum_{i\in\mathcal{D}_j}\mathbf{y}_i\mathbf{y}_i^H\right)\\
&\mathrm{subject\thinspace to}&&
\begin{cases}
\cup_{j=1}^{\NRF}\mathcal{D}_j=\left\{1,\cdots,\Nt\right\}\\
\mathcal{D}_j\cap\mathcal{D}_k=\emptyset,\quad\forall j\ne k,
\end{cases}
\end{aligned}
\end{equation}
where $\mathcal{D}_j$ is the mapping set containing the antenna indices that are mapped to the $j$-th RF chain, and $\lambda_1(\cdot)$ denotes the largest eigenvalue of a matrix. The design problem is a combinatorial problem and thus the optimal
solution can be given by exhaustive search with an extremely huge number of possible mapping strategies, which prevents its practical implementation. Therefore, a greedy algorithm and a modified K-means algorithm were proposed in \cite{doubling}.

\begin{figure}[t]
	\centering
	\includegraphics[height=6.5cm]{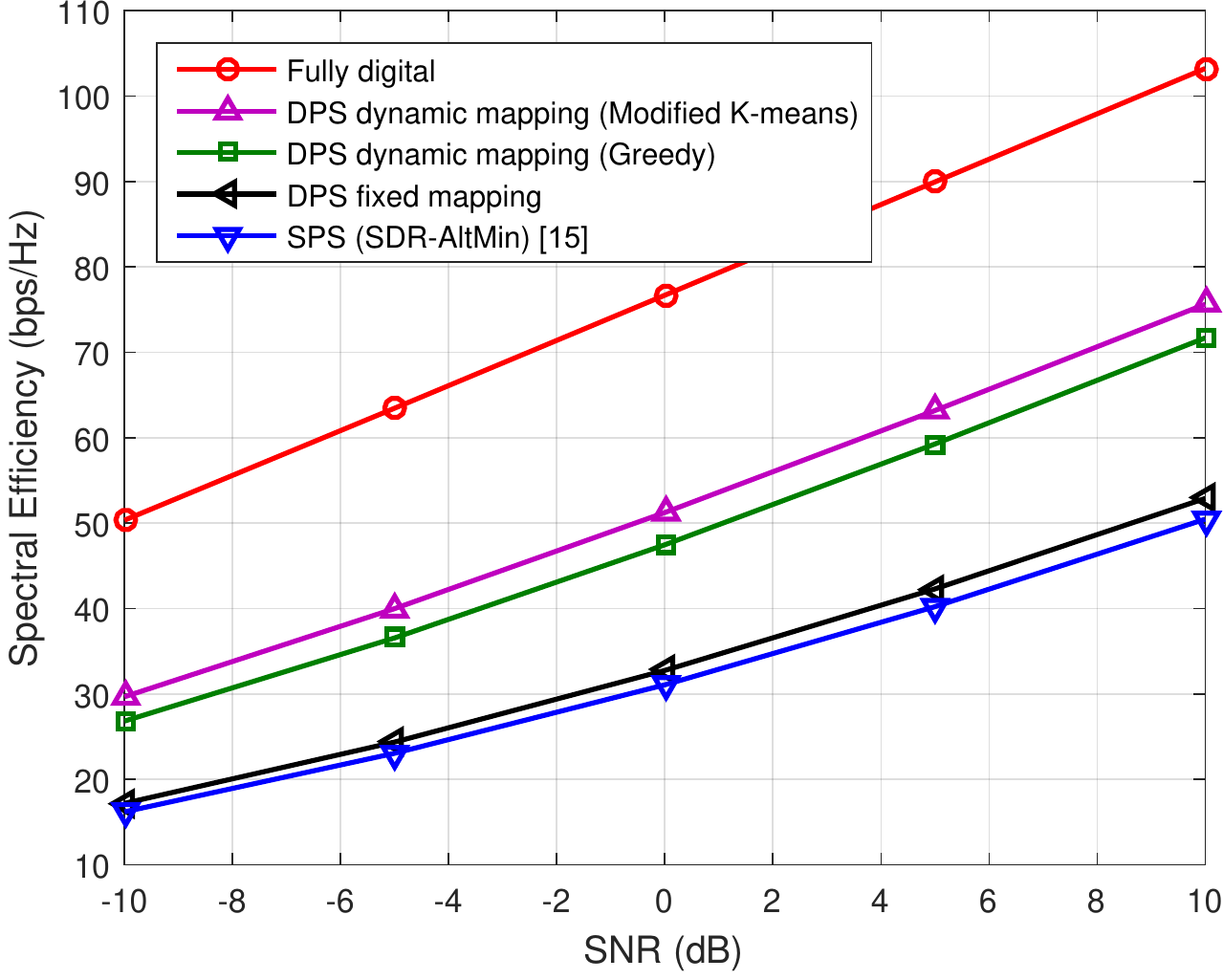}
	\caption{Spectral efficiency achieved by different hybrid beamforming algorithms in the DPS partially-connected structure. There are 4 mobile users and the BS transmits 2 data streams to each user. The BS and user are equipped with 256 and 16 antennas, respectively.}\label{fig6}
\end{figure}

Fig. \ref{fig6} shows the performance of different design approaches
in the DPS partially-connected structure with the minimum
numbers of RF chains, i.e., $\NRF = KNs$.
We see that, due to the sharply reduced number of phase
shifters, the partially-connected structure does entail non-negligible
performance loss compared to the fully digital one.
Furthermore, it shows that
simply doubling the number of phase shifters with the fixed
mapping only has little performance gain over the conventional
SPS implementation \cite{7397861}. Fig. \ref{fig6} demonstrates
that dynamic mapping is able to shrink the gap between the
fixed mapping and the fully digital beamforming by half.

Considering the increased number of phase shifters, the DPS implementation may not be practical for deployment before low-cost low-power phase shifters are available, but it does provide valuable guidelines to design other hybrid beamforming structures.

\begin{enumerate}
	\item With computationally efficient and optimal beamforming algorithms, the DPS fully-connected structure can serve as a performance upper bound for structures that are with higher hardware efficiency. It is a tighter upper bound than the fully digital beamforming, especially when the number of RF chains is small.
	
    \item The computationally efficient algorithm for the DPS fully-connected structure has inspired a highly effective algorithm for the SPS fully-connected structure, which enjoys a low computational complexity and outperforms existing algorithms.

	\item The algorithmic and performance advantages of the DPS implementation are achieved via passing the same signal through more than one phase shifter, which can inspire similar proposals for improvement, as will be discussed in the next subsection.
	
	\item As the beamforming problem becomes a low-rank matrix approximation (eigenvalue) problem for the DPS fully-connected (partially-connected) structure, theoretical analysis, which is intractable for other structures, becomes possible. It will then help to better understand hybrid beamforming systems.
	
\end{enumerate}

\subsection{Fixed Phase Shifter (FPS) Implementation}\label{SecIII-B}
The key weakness of the DPS implementation is the low hardware efficiency. Nevertheless, as discussed above, we can draw valuable lessons for further improvement. The key idea of DPS is to pass the signal out of each RF chain through more than one phase shifter. Specifically, this will help to overcome the non-convex unit modulus constraint for the analog network, and thus significantly simplifies algorithm design. At the same time, it will provide capability to change the amplitudes of elements of the analog beamforming matrix, which helps to improve the spectral efficiency.

Inspired by these insights, a novel analog network implementation, namely the FPS implementation, has been proposed in \cite{8310586}, which allows each signal to pass multiple phase shifters. A key difference compared with previous proposals is that only a small number of phase shifters, with \emph{quantized} and \emph{fixed} phases, are employed. While existing works on hybrid beamforming commonly assumed a large number of phase shifters with unquantized phases, in practice the phase shifters should be discretized with a coarse quantization, and their number should be reduced to a minimum due to cost and power consideration. Thus, the FPS implementation is very promising for practical systems.

\begin{figure}[t]
	\centering
	\includegraphics[height=3cm]{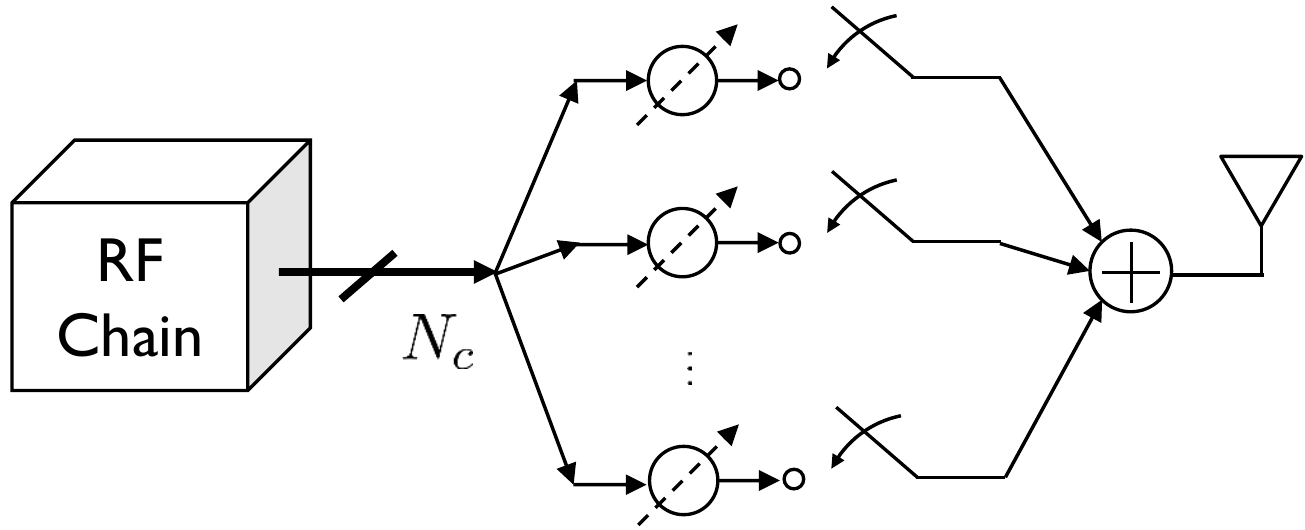}
	\caption{The FPS implementation from an RF chain to a connected antenna.}\label{sub}
\end{figure}

With a small number of fixed phase shifters, the beamformer has limited capability to adapt to the channel states, which will inevitably entail performance loss. To overcome this drawback, a dynamic switch network is cascaded after the fixed phase shifters, as shown in Fig. \ref{sub}. In particular, a total of $N_c$ multichannel ($\NRF$-channel) fixed phase shifters are employed, each of which simultaneously processes the output signals from $\NRF$ RF chains in a parallel manner. In this way, these $N_c$ phase shifters generate $N_c$ signals with different phases for the signal of each RF chain. Inspired by the DPS implementation, a subset of these $N_c$ signals are selected and combined to compose the analog beamforming gain from the RF chain to the antenna. As $N_c$ adaptive switches are needed for each RF chain-antenna pair, in total $\Nt\NRF N_c$ switches are needed for the FPS implementation. The switch network provides dynamic connection from phase shifters to antennas, which is adaptive to channel states. Equipped with a small number of fixed phase shifters and assisted by low-complexity switches, the FPS implementation enjoys hardware complexity comparable to or even lower than the analog beamforming, which needs $\Nt$ phase shifters with adaptive phases.

For beamforming algorithm design, different from other implementations, the analog network of the FPS implementation is essentially to determine the states of different switches, with binary variables, whose formulation is given by
\begin{equation}
\begin{aligned}
&\underset{\mathbf{S},\mathbf{F}_\mathrm{BB}}{\mathrm{minimize}} && \left\Vert\Fopt-\mathbf{SC}\FBB\right\Vert_F^2\\
&\mathrm{subject\thinspace to}&&
\mathbf{S}\in\{0,1\}^{\Nt\times N_c\NRF},
\end{aligned}
\end{equation}
where the switch matrix $\mathbf{S}$ is a binary matrix. The matrix $\mathbf{C}$ stands
for the phase shift operation carried out by the available fixed
phase shifters, given by a block diagonal matrix as
\begin{equation}\label{psmatrix}
\mathbf{C}=\mathrm{blkdiag}\left(\underbrace{\mathbf{c},\mathbf{c},\cdots,\mathbf{c}}_{\NRF}\right),
\end{equation}
where $\mathbf{c}=\frac{1}{\sqrt{N_c}}\left[e^{\jmath\theta_1},e^{\jmath\theta_2},\cdots,e^{\jmath\theta_{N_c}}\right]^T$ is the normalized phase shifter vector containing all $N_c$ fixed phases $\left\{\theta_i\right\}_{i=1}^{N_c}$. Note that although there are $N_c\NRF$ non-zero parameters in matrix $\mathbf{C}$, only $N_c$ phase shifters are required since the phase shifters are with $\NRF$ parallel channels and shared by all RF chain-antenna pairs.
To solve this problem, an efficient AltMin algorithm was proposed in \cite{8310586}. A tight upper bound of the objective function was first derived, based on which {\color{black}closed-form} solutions for both the dynamic switch network and the digital baseband beamformer. Note that we may also develop an FPS partially-connected structure to reduce the number of switches, but it has been found to incur significant performance loss. We will explore a more effective approach to achieve hardware-performance trade-offs in Section \ref{SecIV}.

\subsection{Performance Comparison}
\begin{figure}[t]
	\centering\includegraphics[height=6.5cm]{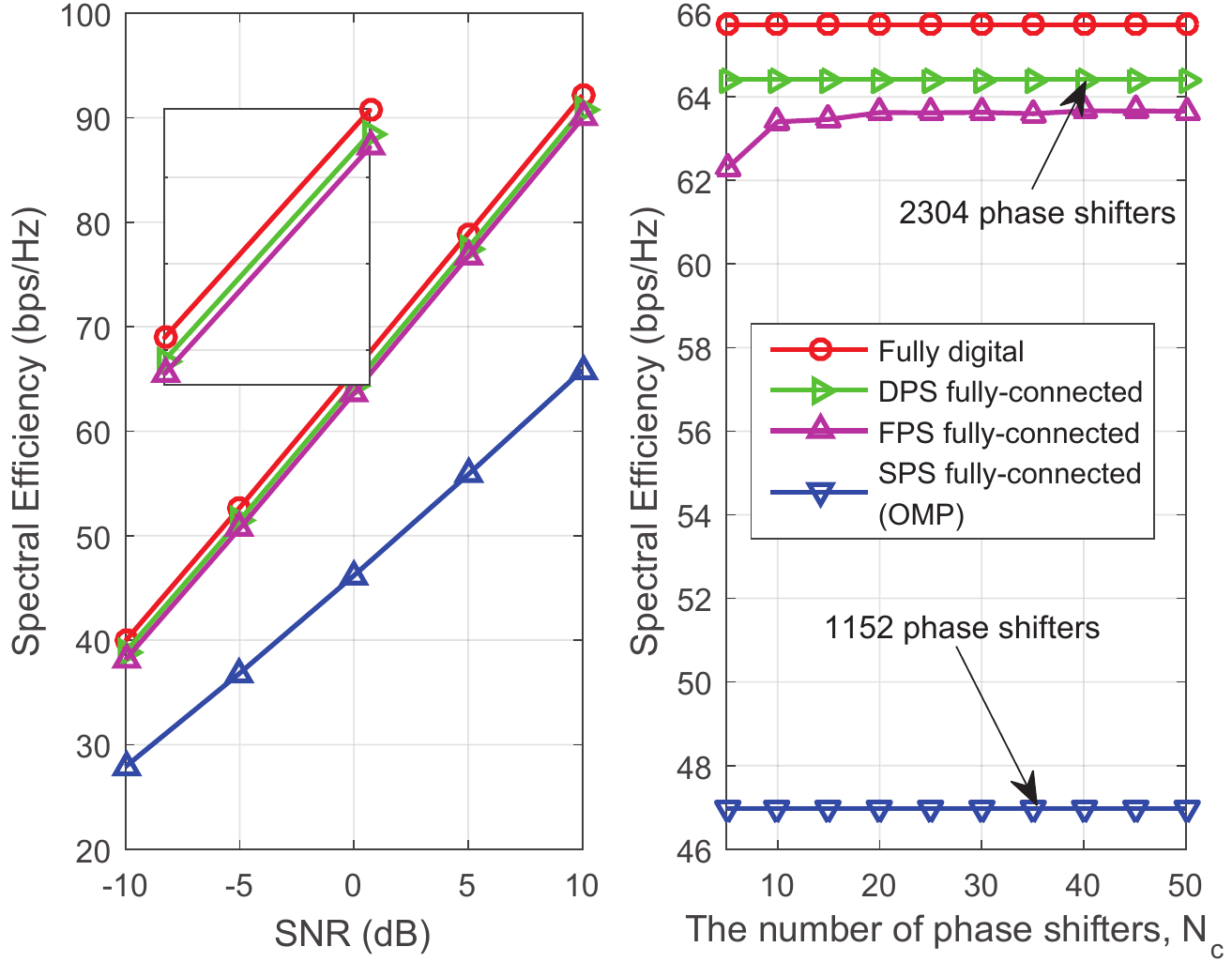}
	\caption{Spectral efficiency of different implementations in a $144\times16$ MIMO-OFDM system with 128 subcarriers. The BS transmits 2 data streams to each of the 4 users on each subcarrier, with 8 RF chains at each node. For the left hand side figure, the number of phase shifters is 30, while we change the number of phase shifters for the FPS implementation in the right hand side figure, with SNR being 0 dB.}
	\label{fig2}
\end{figure}

In Fig. \ref{fig2}, the spectral efficiency of the two presented analog network implementations is evaluated, compared with the fully digital beamforming and the SPS fully-connected structure with the OMP algorithm. As a general multicarrier multiuser system is considered, the MO-AltMin algorithm is inapplicable due to high complexity. It shows that both the DPS and FPS fully-connected structures achieve performance close to the fully digital one. This is quite an astonishing result, given that a single analog network is shared by all the users and subcarriers, and the number of RF chains is only the same as the number of data streams. The poor performance of the SPS implementation is partly due to the sub-optimality of the beamforming algorithm, as the unit modulus constraint in the analog beamforming matrix makes it difficult to develop high-performance low-complexity algorithms.

Remarkably, the FPS fully-connected structure performs closely to the DPS one, though with much fewer phase shifters. As shown in the figure on the right, around 10 fixed phase shifters are sufficient for the FPS implementation, while the SPS and DPS implementations require 1152 and 2304 phase shifters, respectively. This makes the FPS implementation very attractive for practical deployment. Meanwhile, once low-cost high-resolution commercial phase shifters are available, or for cost-insensitive applications, the DPS implementation would be an ideal choice in terms of both the spectral efficiency and computational efficiency.

\begin{table*}[t]
	{\color{black}
	\caption{Comparisons of hardware components in the analog network for different hybrid beamforming structures}
\begin{center}
		\begin{tabular}{|c|c|c|c|c|c|c|c|}
		\hline
		\multicolumn{1}{|c|}{\multirow{2}{*}{\textbf{Implementation}}} & \multicolumn{1}{c|}{\multirow{2}{*}{\textbf{Mapping strategy}}}& \multicolumn{3}{c|}{\textbf{Phase shifter}} & \multicolumn{3}{c|}{\textbf{Other hardware components}} \bigstrut\\\cline{3-8}
		\multicolumn{1}{|c|}{}                                & \multicolumn{1}{c|}{}                                  & {\textbf{Number} } & {\textbf{Type}} & \textbf{Power} & {\textbf{Hardware}} & {\textbf{Number}} & \textbf{Power} \bigstrut[t]\\
		\hhline{|=|=|=|=|=|=|=|=|}
		\multirow{2}[4]{*}{{SPS \cite{6717211,7397861}}} & {Fully-connected} &  $\NRF\Nt$     & \multirow{2}[4]{*}{Adaptive} & \multirow{2}[4]{*}{50 mW} & \multirow{2}[4]{*}{N/A} &    \multirow{2}[4]{*}{N/A}    & \multirow{2}[4]{*}{N/A} \bigstrut\\
		\cline{2-3}          & {Partially-connected} &   $\Nt$    &       &       &       &      &  \bigstrut\\
		\hline
		\multirow{2}[4]{*}{{DPS \cite{doubling}}} & {Fully-connected} &  $2\NRF\Nt$     & \multirow{2}[4]{*}{Adaptive} & \multirow{2}[4]{*}{50 mW} & \multirow{2}[4]{*}{N/A} &    \multirow{2}[4]{*}{N/A}   & \multirow{2}[4]{*}{N/A} \bigstrut\\
		\cline{2-3}       & {Partially-connected} & $2\Nt$      &       &       &       &       &  \bigstrut\\
		\hline
		\multirow{2}[4]{*}{{FPS} \cite{8310586}} & {Fully-connected} & \multirow{2}[4]{*}{$N_c\ll\Nt$}      & Multi-channel & \multirow{2}[4]{*}{20 mW} & \multirow{2}[4]{*}{Switch} &   $N_c\NRF\Nt$    & \multirow{2}[4]{*}{5 mW} \bigstrut\\
		\cline{2-2}\cline{7-7}          & {Group-connected} &       & Fixed      &       &       & $\frac{1}{\eta}N_c\NRF\Nt$       &  \bigstrut\\
		\hline
	\end{tabular}%
\end{center}
\begin{tablenotes}
	\item * The  power consumption of (fixed) phase shifters and switches are typical values in \cite{rappaport2014millimeter} and \cite{5648370}.
\end{tablenotes}
	\label{table2}
}
\end{table*}%
\vspace{1em}
\begin{table*}[t]
	{\color{black}
		\begin{center}
			\caption{Comparisons of computational complexity for different  hybrid beamforming algorithms}
			\begin{tabular}{|c|c|c|c|c|}\hline
				\textbf{Implementation}&\textbf{Mapping strategy}&\textbf{Design algorithm}&\textbf{Computational complexity}&\textbf{Spectral efficiency}\bigstrut\\
				\hhline{|=|=|=|=|=|}
				\multirow{2}[4]{*}{SPS \cite{6717211,7397861}} & Fully-connected &   MO-AltMin \cite{7397861}     &  Extremely high    &  \checkmark\checkmark\checkmark\checkmark\bigstrut\\\cline{2-5}
				& Partially-connected & SDR-AltMin \cite{7397861}     &  $\mathcal{O}\left(N_\mathrm{iter}{\NRF}^3K^3N_s^3F^3\right)$      &\checkmark  \bigstrut\\\hline
				\multirow{2}[4]{*}{DPS \cite{doubling}} & Fully-connected & Matrix decomposition \eqref{eq16}    &  $\mathcal{O}\left(\NRF\Nt KN_s F\right)$    &\checkmark\checkmark\checkmark\checkmark\checkmark\checkmark  \bigstrut\\\cline{2-5}
				& Partially-connected & Modified K-means \cite{doubling}    &    $\mathcal{O}\left(N_\mathrm{
					iter}{\NRF}K^3N_s^3 F^3\right)$    &\checkmark\checkmark  \bigstrut\\\hline
				\multirow{2}[4]{*}{FPS \cite{8310586}} & Fully-connected & \multirow{2}[4]{*}{FPS-AltMin \cite{8310586}}    &  {$\mathcal{O}\Big(N_\mathrm{iter}
					\big(K^2N_s^2\NRF+$  }    &\checkmark\checkmark\checkmark\checkmark\checkmark  \bigstrut\\\cline{2-2}\cline{5-5}
				& Group-connected &  & $N_c\NRF\Nt\log\left(N_c\NRF\Nt\right)\big)
				\Big)$       &\checkmark\checkmark\checkmark  \bigstrut\\\hline
			\end{tabular}\label{table3}
		\end{center}
			
	}
\end{table*}%

\section{A Flexible Mapping Strategy for Hardware-Performance Trade-offs}\label{SecIV}
Among the presented hybrid beamforming structures, the DPS fully-connected structure performs the best in both computational efficiency and spectral efficiency, but with low hardware efficiency. The FPS fully-connected structure achieves a good balance among the three design aspects, but requires a large number of switches. Considering the cost and power consumption of hardware components, especially for mm-wave systems, it is important to further reduce the hardware complexity. Meanwhile, the partially-connected mapping strategy fails to be a good candidate for high hardware efficiency, as it reduces hardware complexity by too much and incurs significant performance loss. Thus, it is highly desirable to have fine granularity when reducing the hardware complexity. In this section, we present a flexible hybrid beamforming mapping strategy, called the \emph{group-connected} mapping, to achieve a better balance between hardware efficiency and spectral efficiency.

As shown in Table I (a), with this new mapping strategy, antennas and RF chains are divided into $\eta$ groups, and signals coming out of {\color{black}each} RF chain group are transmitted via {\color{black}its} corresponding antenna group. The grouping is flexible, and the numbers of RF chains and antennas in different groups can be different. The mapping strategy within each group is the same as the fully-connected mapping. Thus, the analog beamforming matrix $\FRF$ has the block diagonal structure, with each block corresponding to one RF chain-antenna group. It is easy to observe that conventional fully- and partially-connected mapping strategies are special cases of this flexible one:
\begin{itemize}
	\item When $\eta=1$, there is only one RF chain group and one antenna group, and thus we get the fully-connected mapping strategy;
	\item When $\eta=\NRF$, each RF chain group contains a single RF chain, which is connected to a group of antennas, and thus we get the partially-connected mapping strategy.
\end{itemize}

By varying the value of $\eta$, we can easily obtain hybrid beamforming mapping strategies with different hardware complexities. Moreover, we can apply any of the hardware implementations presented in Table I (b) with this group-connected mapping. For the SPS and DPS implementations, the number of phase shifters is $1/\eta$ of the fully-connected one; for the FPS implementation, the number of switches is $1/\eta$ of the one shown in Table I (b), while the number of fixed phase shifter keeps the same.

In terms of beamforming algorithm design, due to the block diagonal structure in $\FRF$, we can decouple the design of each block, for which the problem is similar to the conventional fully-connected mapping. Therefore, we can leverage the rich algorithms presented in the previous two sections for different analog network implementations. In other words, this flexible structure does not introduce any additional difficulty in beamforming algorithm design.

\begin{figure}[t]
	\centering\includegraphics[height=6.5cm]{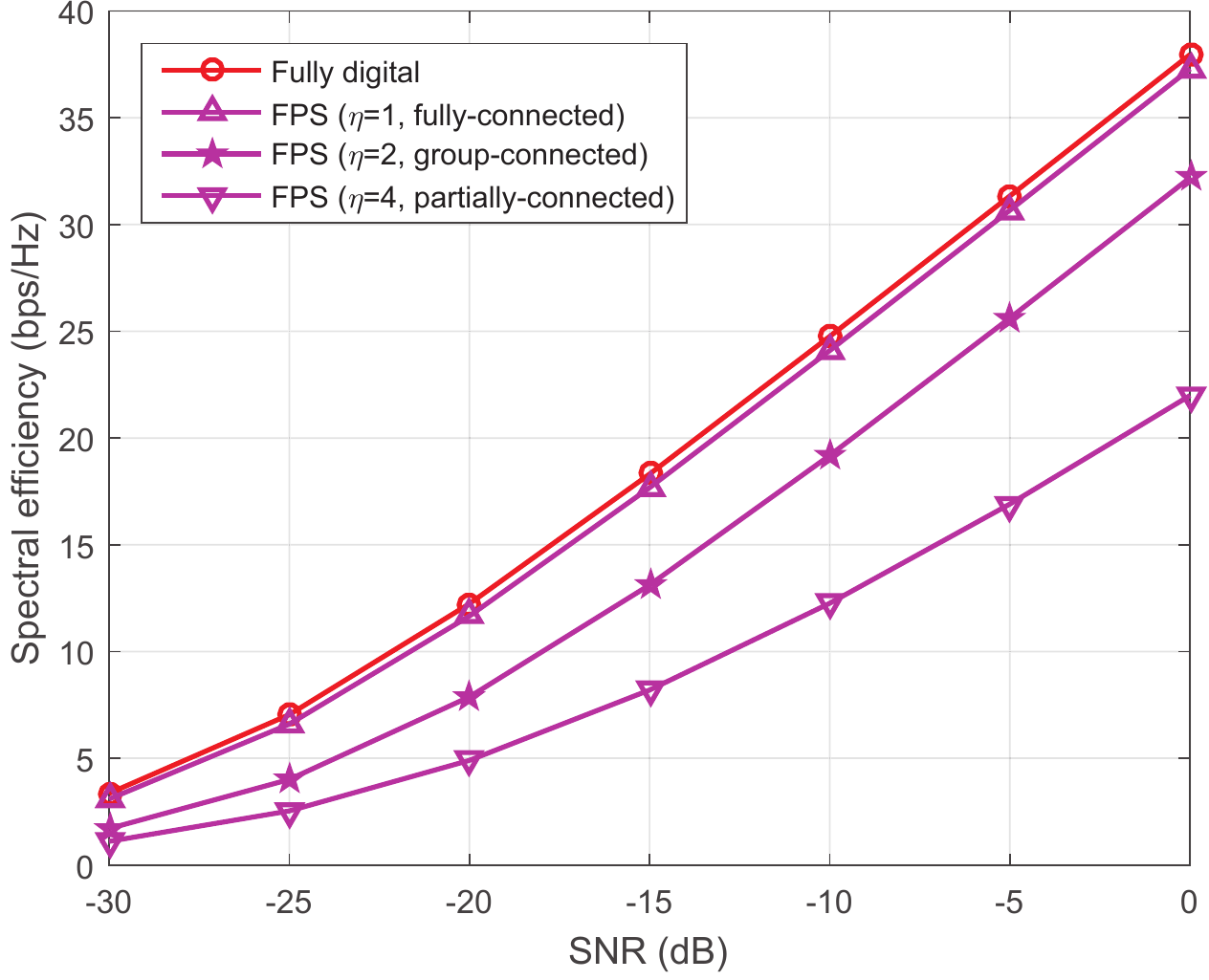}
	\caption{Spectral efficiency of different values of $\eta$ with the FPS group-connected structure. The BS and user are with 256 and 16 antennas. The numbers of RF chains at both the BS and user sides are equal to that of the data streams, which is 4.}
	\label{fig3}
\end{figure}

In Fig. \ref{fig3}, we compare spectral efficiency of the FPS group-connected structure with different values of $\eta$. Other implementations have the same trend. It shows that varying the value of $\eta$ helps to effectively balance the hardware complexity and spectral efficiency. To summarize, this new mapping strategy enjoys the following three desirable properties:

\begin{enumerate}
	\item	It provides a flexible way to trade off performance against hardware complexity;
	
	\item	It is compatible with different analog network implementations;
	
	\item	The hybrid beamformer can be effectively designed by leveraging existing algorithms.
\end{enumerate}

Therefore, this mapping strategy, especially with the FPS implementation, stands out as a promising candidate to support hybrid beamforming in 5G and beyond mm-wave systems. 
{\color{black}The hardware components in the analog network and design algorithms for different hybrid beamforming structures are compared in Tables \ref{table2} and \ref{table3}, respectively.}

\section{Conclusions and Future Directions}\label{SecV}
\begin{figure}[t]
	\centering\includegraphics[height=8cm]{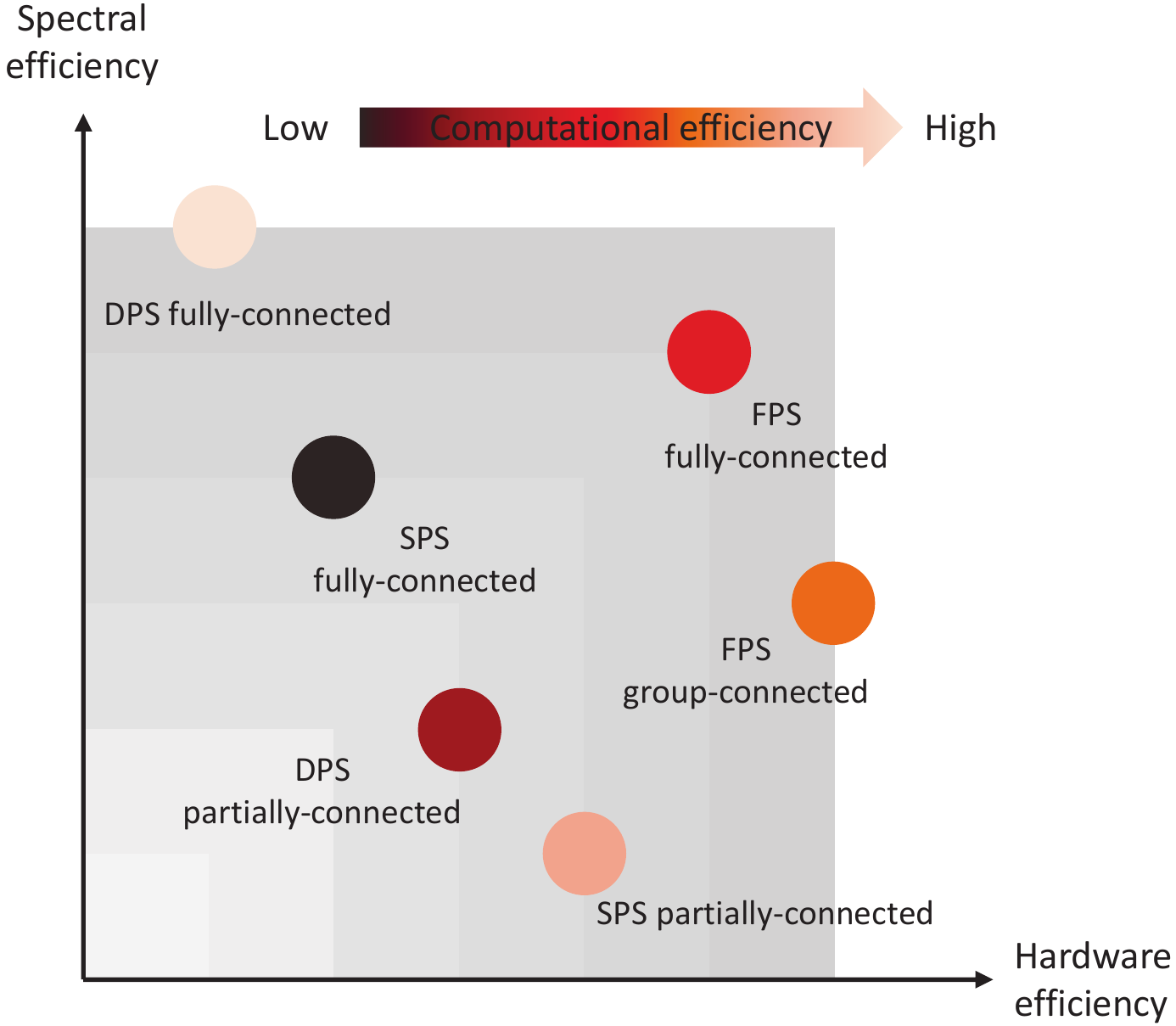}
	\caption{Qualitative comparison of different structures, in hardware efficiency, computation efficiency of the typical algorithm, and achievable spectral efficiency.}
	\label{fig4}
\end{figure}

In this paper, we presented several proposals of hybrid beamforming structures in mm-wave systems, focusing on three key aspects: hardware efficiency, spectral efficiency, and computational efficiency.  Through a systematic comparison, important design insights were revealed. In particular, it was shown that hardware implementation significantly affects the algorithm design and achievable spectral efficiency. With a suitable structure, hybrid beamforming can approach the performance of the fully digital one with low hardware complexity. For example, it is sufficient to have RF chains comparable to the number of data streams, and a small number ($\sim$10) of fixed phase shifters are sufficient with the FPS implementation. Furthermore, a flexible structure was proposed to balance hardware efficiency and spectral efficiency. A qualitative comparison of different structures is shown in Fig. \ref{fig4}. Overall, the FPS group-connected structure stands out as a promising candidate for hybrid beamforming in 5G and beyond mm-wave systems. Once low-cost phase shifters are available, the DPS implementation will also be attractive.

To achieve the full success of hybrid beamforming, more works will be needed, and the {\color{black}followings} are some potential future research directions.
\begin{itemize}
\item \textbf{CSI acquisition for hybrid beamforming}: Perfect channel state information (CSI) was assumed in the discussion of this paper, and acquiring large-scale CSI with reduced RF chains is a challenging problem, with some prior studies in \cite{7037320,7178503,6847111,7454701}. Different training methods may be needed for different hybrid beamforming structures {\color{black}\cite{7370753,7961152}}. The presented results also shed light on hybrid beamforming design during the training stage, which is critical to overcome the low SNR during training. {\color{black}In addition, codebook design for channel estimation is of another  particular interest in mm-wave MIMO systems \cite{7845674,7604098,7990158}.}


\item \textbf{Deep learning for efficient hybrid beamforming}: It is highly desirable to further reduce the computational complexity of hybrid beamforming algorithms. Recently, deep learning has been applied to develop efficient algorithms for large-scale optimization problems in wireless networks \cite{8714026,8052521,SpatialDL,shen2018lora}, including hybrid beamforming \cite{7997065,8322248,lin2019beamforming}. While these initial attempts have demonstrated the effectiveness of deep learning-based methods, more investigation will be needed, from both practical and theoretical perspectives.

\item \textbf{Finite-precision ADCs}: While the focus in this paper is on the analog network, there are still some gaps to fill in the digital domain. In particular, the quantization effect of ADCs cannot be ignored. How to extend the presented hybrid beamforming structures to systems with low-resolution ADCs deserves delicate investigation, and some previous studies can be found in \cite{7876856,8320852,8465975,7961157}.

\item \textbf{Algorithm-hardware co-design}: To effectively design the increasingly complex wireless systems, collaboration among the hardware and algorithm domains will be needed.  Hardware-algorithm co-design will play an important role in 5G and beyond systems \cite{8808168}. The target is to develop hardware-efficient transceiver structures that are also algorithm friendly. The FPS implementation can be regarded as a preliminary attempt of such design approaches in mm-wave systems.

\item \textbf{Hybrid beamforming in networks}: From the network perspective, while mm-wave networks with analog beamforming have been extensively analyzed \cite{7913628,7105406,6932503}, the effect of adopting hybrid beamforming has not been fully unraveled. Indeed, hybrid beamforming will result in more intricate signal and interference distributions, which should be carefully investigated.
\end{itemize}

\bibliographystyle{IEEEtran}
\bibliography{open}

\begin{IEEEbiography}
	[{\includegraphics[width=1in,height=1.25in,clip,keepaspectratio]{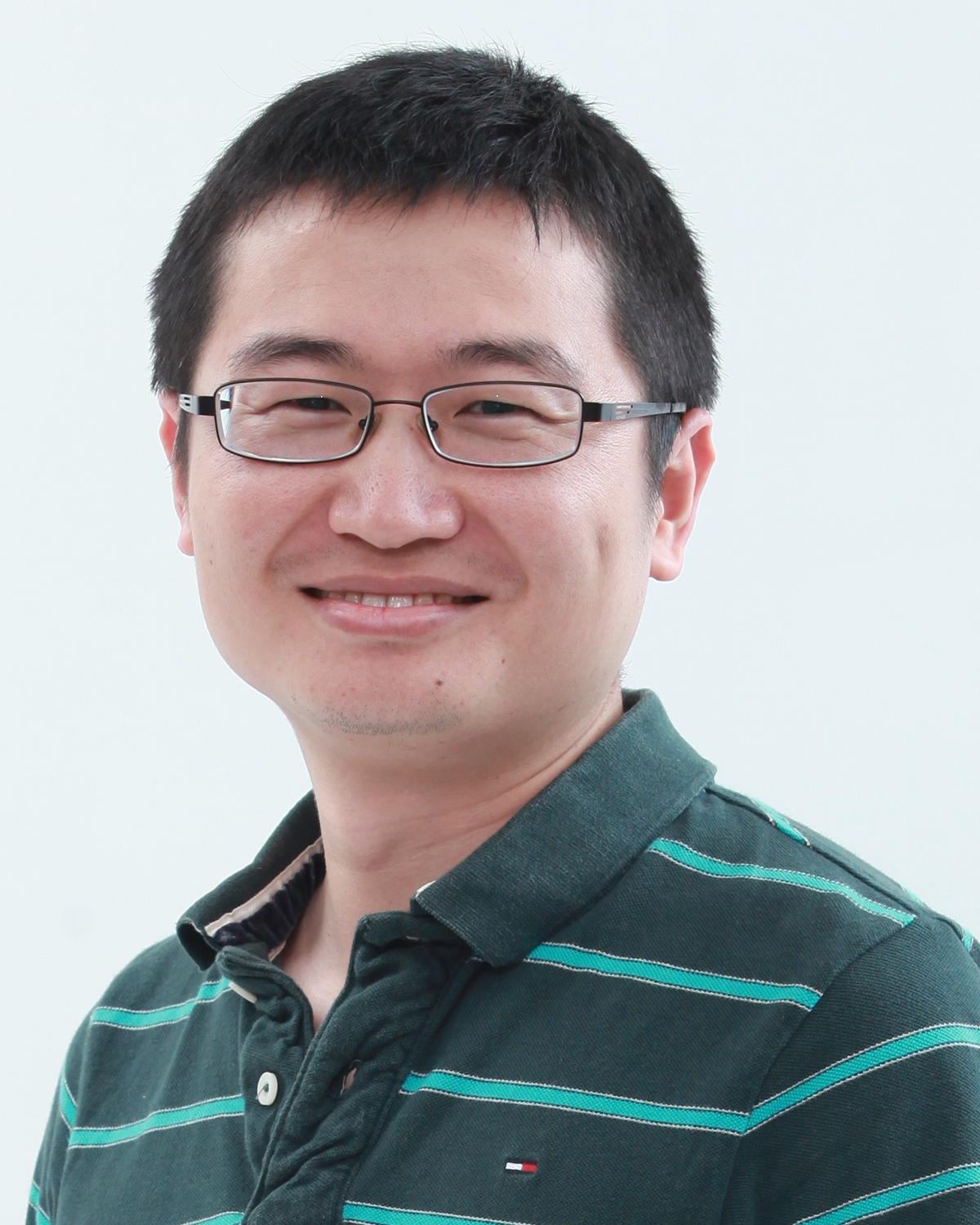}}]{Jun Zhang}
	(S'06-M'10-SM'15) received the
	B.Eng. degree in Electronic Engineering from
	the University of Science and Technology of
	China in 2004, the M.Phil. degree in Information
	Engineering from the Chinese University of Hong
	Kong in 2006, and the Ph.D. degree in Electrical
	and Computer Engineering from the University
	of Texas at Austin in 2009. He is an Assistant
	Professor in the Department of Electronic and
	Information Engineering at the Hong Kong Polytechnic University (PolyU). His research interests
	include wireless communications and networking, mobile edge computing
	and edge learning, distributed learning and optimization, and big data
	analytics.
	
	Dr. Zhang co-authored the books \emph{Fundamentals of LTE} (Prentice-Hall,
	2010), and \emph{Stochastic Geometry Analysis of Multi-Antenna Wireless
	Networks} (Springer, 2019). He is a co-recipient of the 2019 IEEE Communications Society \& Information Theory Society Joint Paper Award, the
	2016 Marconi Prize Paper Award in Wireless Communications (the best
	paper award of \textsc{IEEE Transactions on Wireless Communications}), and
	the 2014 Best Paper Award for the \emph{EURASIP Journal on Advances in
	Signal Processing}. Two papers he co-authored received the Young Author
	Best Paper Award of the IEEE Signal Processing Society in 2016 and
	2018, respectively. He also received the 2016 IEEE ComSoc Asia-Pacific
	Best Young Researcher Award. He is an Editor of \textsc{IEEE Transactions
	on Wireless Communications}, \textsc{IEEE Transactions
	on Communications}, and \textsc{Journal of Communications and
	Information Networks}, and an IEEE senior member.
\end{IEEEbiography}

\begin{IEEEbiography}
	[{\includegraphics[width=1in,height=1.25in,clip,keepaspectratio]{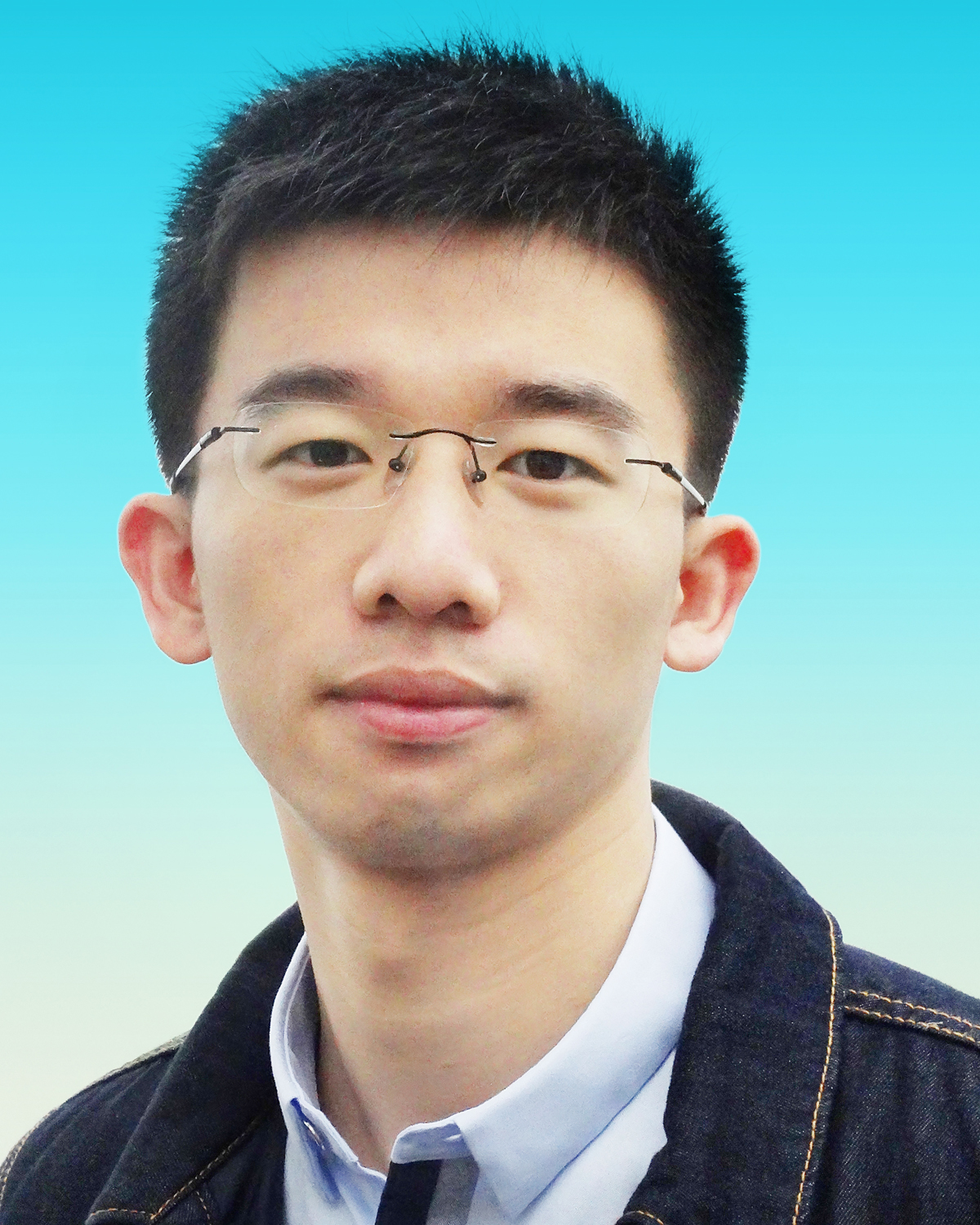}}]{Xianghao Yu}
	(S'15--M'19) received the B.Eng. degree in Information Engineering from Southeast University (SEU), Nanjing, China, in 2014, and the Ph.D. degree in Electronic and Computer Engineering from the Hong Kong University of Science and Technology (HKUST), Kowloon, Hong Kong, in 2018. He is currently a Humboldt Postdoctoral Research Fellow at the Institute for Digital Communications, Friedrich-Alexander-University Erlangen-Nurnberg (FAU), Erlangen, Germany. His research interests include millimeter wave communications, intelligent reflecting surface assisted communications, MIMO systems, mathematical optimization, and stochastic geometry. Dr. Yu is a co-author of the book \emph{Stochastic Geometry Analysis of Multi-Antenna Wireless Networks} (Springer, 2019). He was the recipient of the IEEE GLOBECOM 2017 Best Paper Award, the 2018 IEEE Signal Processing Society Young Author Best Paper Award, and the IEEE GLOBECOM 2019 Best Paper Award.
\end{IEEEbiography}

\begin{IEEEbiography}
	[{\includegraphics[width=1in,height=1.25in,clip,keepaspectratio]{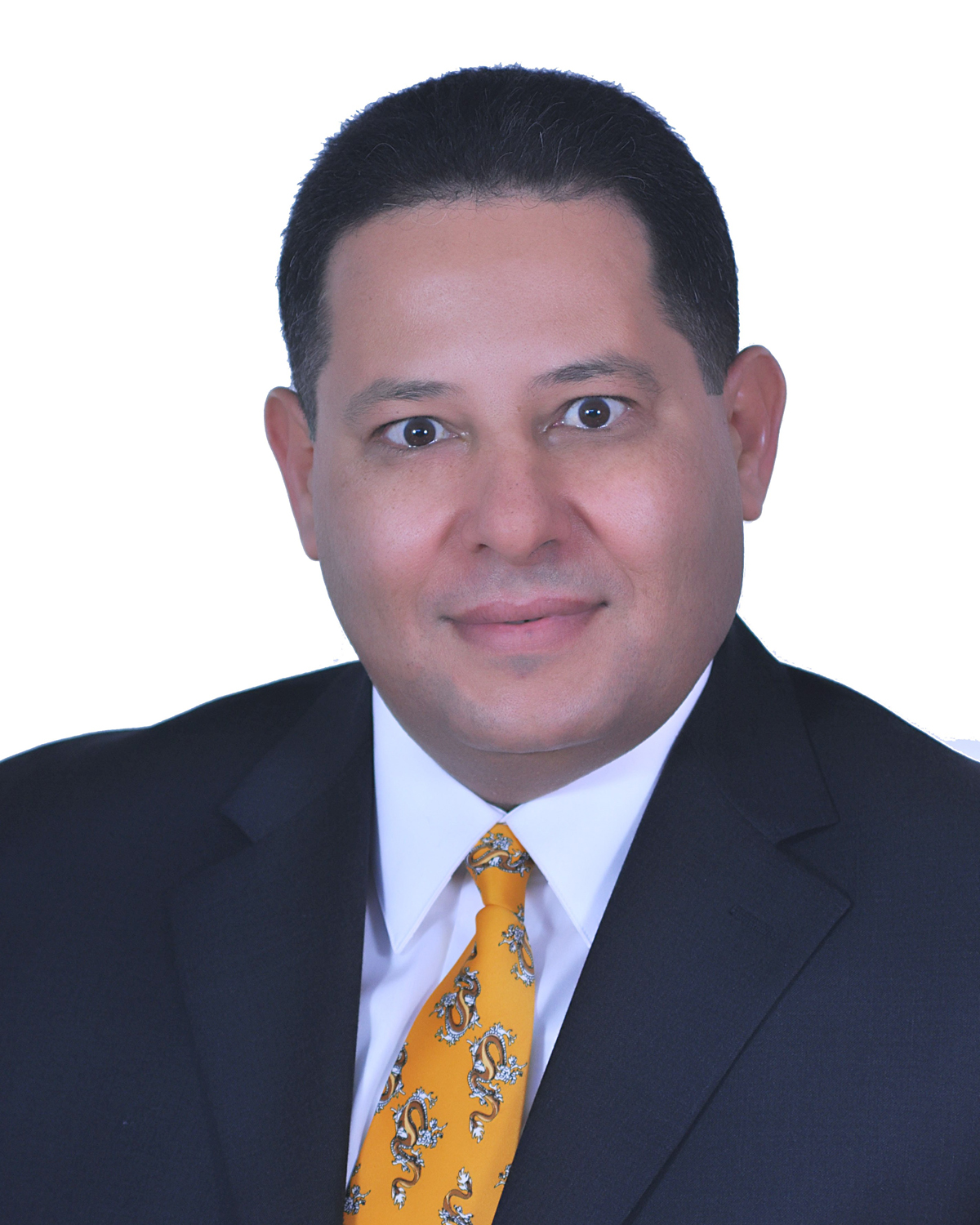}}]{Khaled B. Letaief}
	(S'85--M'86--SM'97--F'03) received the BS degree \emph{with distinction} in Electrical Engineering from Purdue University at West Lafayette, Indiana, USA, in December 1984. He received the MS and Ph.D. Degrees in Electrical Engineering from Purdue University, in August 1986, and May 1990, respectively.
	
	From 1990 to 1993, he was a faculty member at the University of Melbourne, Australia.  He has been with the Hong Kong University of Science \& Technology since 1993 where he is known as one of HKUST's most distinguished professors for his boundless energy, collegial nature, dedication, and excellence in research, education, and service.  While at HKUST, he has held numerous administrative positions, including  Head of the Electronic and Computer Engineering department, and Director of the Hong Kong Telecom Institute of Information Technology.
	
	While at HKUST he has also served as Chair Professor and Dean of HKUST School of Engineering. Under his leadership, the School of Engineering has not only transformed its education and scope and produced very high caliber scholarship, it has also actively pursued knowledge transfer and societal engagement in broad contexts. It has also dazzled in international rankings (rising from \#26 in 2009 to \#14 in the world in 2015 according to QS World University Rankings).
	From 2015 to 2018, he joined HBKU as Provost to help establish a research-intensive university in Qatar in partnership with strategic partners that include Northwestern University, Carnegie Mellon University, Cornell, and Texas A\&M.
	
	Dr. Letaief is an internationally recognized leader in wireless communications and networks with research interest in artificial intelligence, big data analytics systems, mobile cloud and edge computing, tactile Internet, 5G systems and beyond.  In these areas, he has over 620 journal and conference papers and given keynote talks as well as courses all over the world.  He also has 15 patents, including 11 US patents.
	
	Dr. Letaief served as consultants for different organizations including Huawei, ASTRI, ZTE, Nortel, PricewaterhouseCoopers, and Motorola.  He is the founding Editor-in-Chief of the \textsc{IEEE Transactions on Wireless Communications} and has served on the editorial board of other prestigious journals including the \textsc{IEEE Journal on Selected Areas in Communications}-Wireless Series (as Editor-in-Chief).  He has been involved in organizing a number of flagship international conferences and events. 
	
	In addition to his active research and professional activities, Professor Letaief has been a dedicated teacher committed to excellence in teaching and scholarship.  He received the Mangoon Teaching Award from Purdue University in 1990; the Teaching Excellence Appreciation Award by the School of Engineering at HKUST (4 times); and the Michael G. Gale Medal for Distinguished Teaching (\emph{Highest university-wide teaching award} and only one recipient/year is honored for his/her contributions).
	
	He is also the recipient of many other distinguished awards and honors including 2019 Distinguished Research Excellence Award by HKUST School of Engineering (Highest research award and only one recipient/3 years is honored for his/her contributions); 2019 IEEE Communications Society and Information Theory Society Joint Paper Award; 2018 IEEE Signal Processing Society Young Author Best Paper Award; 2017 IEEE Cognitive Networks Technical Committee Publication Award; 2016 IEEE Signal Processing Society Young Author Best Paper Award; 2016 IEEE Marconi Prize Paper Award in Wireless Communications; 2011 IEEE Wireless Communications Technical Committee Recognition Award; 2011 IEEE Communications Society Harold Sobol Award; 2010 Purdue University Outstanding Electrical and Computer Engineer Award; 2009 IEEE Marconi Prize Award in Wireless Communications; 2007 IEEE Communications Society Joseph LoCicero Publications Exemplary Award; and over 15 IEEE Best Paper Awards.
	
	Dr. Letaief is well recognized for his dedicated service to professional societies and in particular IEEE where he has served in many leadership positions.  These include Treasurer of IEEE Communications Society, IEEE Communications Society Vice-President for Conferences, Chair of IEEE Committee on Wireless Communications, elected member of IEEE Product Services and Publications Board, and IEEE Communications Society Vice-President for Technical Activities. He is currently President of the IEEE Communications Society, the world's leading organization for communications professionals with headquarter in New York City and members in 162 countries.
	
	Dr. Letaief is a Fellow of IEEE and a Fellow of HKIE.  He is also recognized by Thomson Reuters as an \emph{ISI Highly Cited Researcher}.
\end{IEEEbiography}

\end{document}